\documentclass[12pt,a4paper]{article}
\pdfoutput=1
\usepackage{jheppubx}
\usepackage{amsmath,amsfonts,amssymb,amsthm,graphicx,color}
\usepackage{verbatim}
\usepackage{enumerate}
\numberwithin{equation}{section}
\usepackage{bbm,bm}
\usepackage{braket,mathrsfs}
\usepackage{xcolor}
\usepackage{framed}
\usepackage{tikz}
\usetikzlibrary{decorations.pathmorphing,arrows.meta,bending,snakes}
\usepackage{caption}
\usepackage{subcaption}
\usepackage{dsfont}

\definecolor{darkgreen}{rgb}{0,0.4,0}
\definecolor{darkred}{rgb}{0.4,0,0}
\definecolor{darkblue}{rgb}{0,0,0.4}

\usepackage[framemethod=TikZ]{mdframed}
\definecolor{darkcerulean}{rgb}{0.03, 0.27, 0.49}
	
\newcounter{ex}[section]
\newenvironment{ex}[3][]{%
\refstepcounter{ex}%
\ifstrempty{#2}%
{\mdfsetup{%
frametitle={%
\tikz[baseline=(current bounding box.east),outer sep=0pt]
\node[anchor=east,rectangle,fill=white]
{\strut };}}
}%
{\mdfsetup{%
frametitle={%
\tikz[baseline=(current bounding box.east),outer sep=0pt]
\node[anchor=east,rectangle,fill=darkcerulean!12]
{ #1:{\normalfont ~#2} };}}%
}%
\mdfsetup{innertopmargin=9pt,linecolor=darkcerulean,%
linewidth=1.2pt,topline=true,%
frametitleaboveskip=\dimexpr-\ht\strutbox\relax
}
\begin{mdframed}[]\relax%
\label{#3}}{\end{mdframed}}

\begin{document}
\title{Bounds on the density of states and the spectral gap in CFT$_{2}$}
\author[q]{Shouvik Ganguly,}

\affiliation[q]{Department of Electrical and Computer Engineering, University of California, San Diego\\
La Jolla, CA 92093, USA}

\author[\ \!\tilde{q},\tilde{Q}]{Sridip Pal}

\affiliation[\tilde{q}]{Department of Physics, 
University of California, San Diego,
La Jolla, CA 92093, USA} 
\affiliation[\tilde{Q}]{School of Natural Sciences, 
Institute for Advanced Study
Princeton, NJ 08540, USA}

\emailAdd{shgangul@eng.ucsd.edu}
\emailAdd{sridip@ias.edu}

\abstract{We improve the recently discovered upper and lower bounds on the $O(1)$ correction to the Cardy formula for the density of states integrated over an energy window (of width $2\delta$), centered at high energy in 2 dimensional unitary and modular invariant conformal field theory. We prove optimality of the lower bound for $\delta\to 1^{-}$.\!~We prove a conjectured upper bound on the asymptotic gap between two consecutive Virasoro primaries for a central charge greater than $1,$ demonstrating it to be $1$. Furthermore, a systematic method is provided to establish a limit on how tight the bound on the $O(1)$ correction to the Cardy formula can be made using bandlimited functions. The techniques and the functions used here are of generic importance whenever the Tauberian theorems are used to estimate some physical quantities.}

\maketitle
%



\section{The premise and the results}
Modular invariance is a powerful constraint on the data of $2$D  conformal field theory (CFT).  It relates the low temperature data to the high temperature data. For example, using the fact that the low temperature behavior of the $2$D CFT partition function is universal and controlled by a single parameter $c$, the central charge of the CFT, we can deduce the universal behavior of the partition function, $Z(\beta)$ at high temperature ($\beta\to0$). Thus we can derive the asymptotic behavior of  the density of states, which controls the high temperature behavior of a $2$D CFT \cite{cardy1986operator}. Schematically, we have
\begin{equation}\label{schematicCardy}
\rho(\Delta\to\infty)\simeq \text{Inverse Laplace} \left[Z(\beta\to 0)\right]=\text{Inverse Laplace} \left[Z\left(\frac{4\pi^2}{\beta}\to \infty\right)\right]\,,
\end{equation}
where $\rho(\Delta)$ is the density of states and the modular invariance tells us $Z\left(\frac{4\pi^2}{\beta}\right)=Z(\beta)$. Similar ideas can be extended to one point functions as well, where the low temperature behavior is controlled by the low lying spectra and three point coefficients \cite{KM,dattadaspal}. Yet another remarkable implication of the modular invariance of the partition function is  the existence of infinite Virasoro primaries for CFT with $c>1$. Significant progress has been made in recent years toward exploiting the modular invariance to derive universal results in $2$D CFT under the umbrella of modular bootstrap \cite{Hellerman:2009bu,KM,dattadaspal,HKS,dyer,Das:2017cnv,Collier:2016cls,Collier:2018exn,Cho:2017fzo,Bae:2017kcl}.

The recent entry in the toolkit of modular bootstrap program is Tauberian theorems. The usefulness of Tauberian theorems in the context of CFT is pointed out in \cite{pappadopulo2012operator}; subsequently, its importance has been emphasized in Appendix C of \cite{dattadaspal}, where the authors used Ingham's theorem \cite{ingham1941tauberian}. The fact that going out to the complex plane while using Tauberian theorems would provide extra mileage in controlling the correction terms in various asymptotic quantities of CFT, has been pointed out in \cite{mukhametzhanov2018analytic}. In particular, the use of \cite{subhankulov1976tauberian} turned out to be extremely useful in this context. Recently, using complex Tauberian theorem, Mukhametzhanov and Zhiboedov \cite{Baur} explored the regime of validity, as well as corrections, to the Cardy formula, obtained by using \eqref{schematicCardy}. In particular, they  investigated the entropy $S_{\delta}$ associated with a particular energy window of width $\delta$ around a peak value $\Delta$, which is allowed to go to infinity, and found
\begin{align}\label{eq:master}
S_{\delta}=\log \left(\int_{\Delta-\delta}^{\Delta+\delta} d\Delta^\prime\ \rho(\Delta^\prime)\right)\underset{ \Delta\to \infty}{\simeq} 2\pi \sqrt{\frac{c\Delta}{3}}+\frac{1}{4}\log\left(\frac{c\delta^4}{3\Delta^3}\right)+s(\delta,\Delta)\,,
\end{align}
where $\rho(\Delta)$ is the density of states, given by a sum of Dirac delta functions peaked at the positions of the operator dimensions. It is shown in \cite{Baur} that for $O(1)$ energy width, the $O(1)$ correction $s(\delta,\Delta)$ is bounded from above and below by two $\Delta$ independent functions $s_{\pm}(\delta)$:\begin{align}\label{spm}
\delta=O(1):\, \ s_{-}(\delta)\le s(\delta,\Delta)\le s_+(\delta)
\end{align}

In particular, \cite{Baur} showed that $O(1)$ is the optimal order. One cannot theoretically obtain a universal correction, which is further suppressed compared to the $O(1)$ number without assuming anything beyond unitarity and modularity. This approach can be contrasted to the one taken in \cite{10.2307/2371313} where a convergent Rademacher sum is written down for the Fourier coefficient of Klein invariant function; in fact, such convergent sums can be derived for any CFT with holomorphic modular invariant partition function. The holomorphicity is an extra input in such scenario\footnote{An attempt to extend the Rademacher sum for non-holomorphic CFTs is explored in \cite{Alday:2019vdr}. Nonetheless, the motivation there is to come up with a partition function for pure gravity.}. Since $O(1)$ is the optimal order without any extra input, it is meaningful to improve upon the numerical value of the $O(1)$ correction and look for the optimal bound i.e.\!~to minimize $s_+-s_-$, which is equivalent to maximizing $s_-$ and minimizing $s_+$.\!~One of the purposes of the current note is precisely so, to improve the bounds $s_{\pm}$ and possibly reach optimality. From now on, we always stay at $O(1)$ and improvement/optimality of the bound refers to the numerical value of $s_{\pm}$. To prove optimality, one needs a CFT spectra saturating \eqref{spm}. It turns out, as we will explain shortly, that the numbers $s_{\pm}$ can be derived using functions with bounded Fourier support a.k.a bandlimited functions. We provide a systematic way to estimate how tight the numerical value of the bounds can be made using these bandlimited functions. We find that there is a limit on how much we can push up $s_-$ and push down $s_+$ if we use bandlimited functions. We call this bound on bounds. If the bound on bounds coincides with the achievable bound, that's the best that can be achieved by using bandlimited functions. In general, the actual optimal bound can be different from the best one achieved by using bandlimited functions. Nonetheless, we will show that for a specific value of the energy width, it is possible to achieve the optimal lower bound using bandlimited functions only.\\

$\bullet$ \textbf {Results:}
We prove the conjectured upper bound on the asymptotic gap between Virasoro primaries, which turns out to be $1$. This gap is optimal since for the Monster CFT, the gap is precisely $1$. This provides a universal bound on how sparse a CFT spectrum can be asymptotically. In particular, this rules out the possibility of having a primary spectrum with Hadamard gap: a spectrum $\{\Delta_k\}$ is said to have Hadamard gap if there exists a fixed $\lambda>1$ such that $\frac{\Delta_{k+1}}{\Delta_k} > \lambda$ for all $k$. We also remark that we have found two different ways to prove the optimal bound on the gap, one of which is related to the problem of finding the optimal lower bound $s_-$.

We unravel a curious connection between the sphere packing problem and the problem of finding the optimal lower bound using bandlimited function. This connection is interesting in its own right and is explored in \S\ref{sec:spack} and provides an upper bound on the best possible lower bound for $\delta\leq 1$ using bandlimited functions. For $\delta=1$, the upper bound on the lower bound becomes achievable and thus becomes the best possible bound using bandlimited functions. It turns out that one can do better and prove optimality of the lower bound in $\delta\to 1^{-}$ limit showing that Monster CFT saturates the lower bound.

We improve the bound on the order one correction\footnote{\textit{Update: The optimal/best possible bounds has been found in \cite{Mukhametzhanov:2020swe}. The paper \cite{Mukhametzhanov:2020swe} appeared long after this paper appeared in the arXiv, in particular, while this paper was under review. The results in \S\ref{der} and \S\ref{bonb} have been made sharper in \cite{Mukhametzhanov:2020swe} }}. Let us briefly recall the origin of $s_{\pm}(\delta)$ \cite{Baur}. The actual density of states is a distribution; thus, the naive inverse Laplace transform of the partition function does not make much sense:
$$\rho(\Delta) \underset{?}{=} \int_{-\infty}^{\infty}\text{d}t\ Z(\beta+\imath t) e^{(\beta+\imath t) (\Delta-c/12)}\,,$$
we control this by cutting off the $t$ integral. This is achieved by introducing two functions $\widehat{\phi}_{\pm}(t)$ which have finite bounded support, in particular, the support is a subset of $[-2\pi,2\pi]$. We convolve the partition function with $\widehat{\phi}_{\pm}(t)$. In the $\Delta'$ domain, this induces a smearing over $\rho(\Delta')$ with the Fourier transform of $\widehat{\phi}_{\pm}(t)$ (which we denote as $\phi_{\pm}(\Delta')$). Furthermore, we choose these functions in a manner so that $\phi_{\pm}$ approximates $\Theta$, the indicator function of the interval $[\Delta-\delta,\Delta+\delta]$ from above and below respectively:
$$\phi_-(\Delta')\leq \Theta\left(\Delta'\in[\Delta-\delta,\Delta+\delta]\right)\leq \phi_+(\Delta')\,.$$ The bound on order one correction $s_{\pm}$ can ultimately be related to $\widehat{\phi}_{\pm}(0)$. Often we write $c_{\pm}\equiv \exp\left[s_{\pm}\right]$. In terms of $\widehat{\phi}_{\pm}(0)$ we have $$c_{\pm}\equiv \exp\left[s_{\pm}\right]=\frac{\pi}{\delta}\widehat{\phi}_{\pm}(0)\,,$$ which we will be interchangeably using. With this variable, we can rewrite eq.~\eqref{eq:master} as
\begin{equation}
\begin{aligned}
&2\pi\widehat{\phi}_{-}(0)\left(\frac{c}{48\Delta^3}\right)^{1/4}\exp\left[2\pi\sqrt{\frac{c\Delta}{3}}\right]\\
&\leq \int_{\Delta-\delta}^{\Delta+\delta}d\Delta^\prime\ \rho(\Delta^\prime) 
\le\\
& 2\pi\widehat{\phi}_{+}(0)\left(\frac{c}{48\Delta^3}\right)^{1/4}\exp\left[2\pi\sqrt{\frac{c\Delta}{3}}\right]\,.
\end{aligned}
\end{equation}

\begin{figure}[h!]
 \centering
    \begin{subfigure}[b]{0.45\textwidth}
        \centering
\includegraphics[scale=0.7]{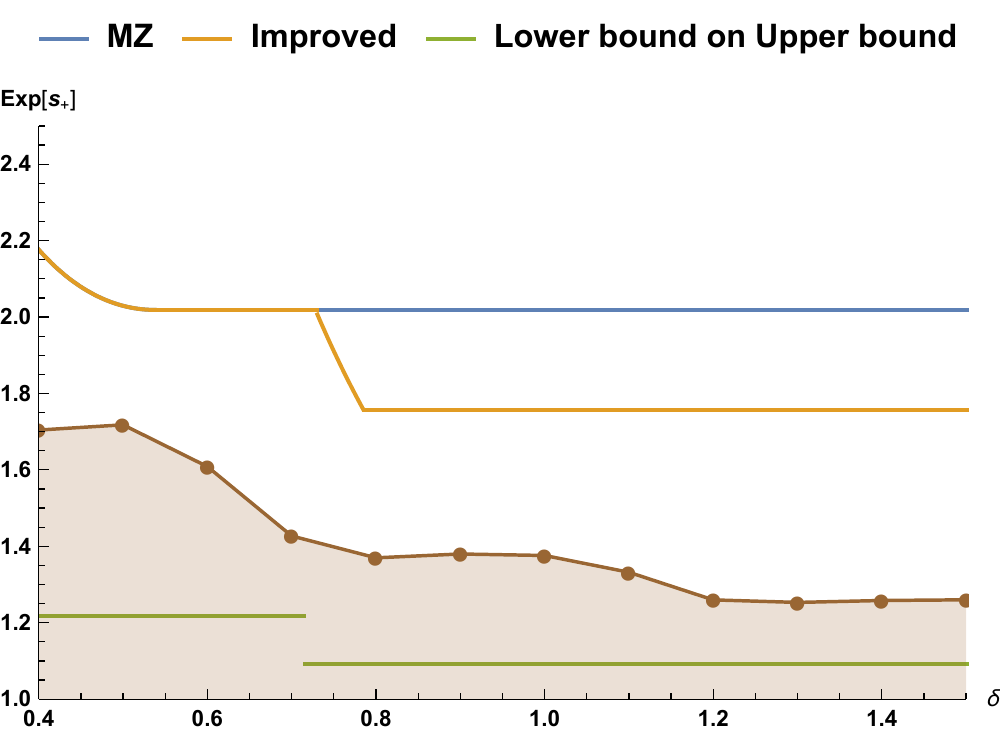}
\end{subfigure}
~
 \begin{subfigure}[b]{0.45\textwidth}
        \centering
\includegraphics[scale=0.7]{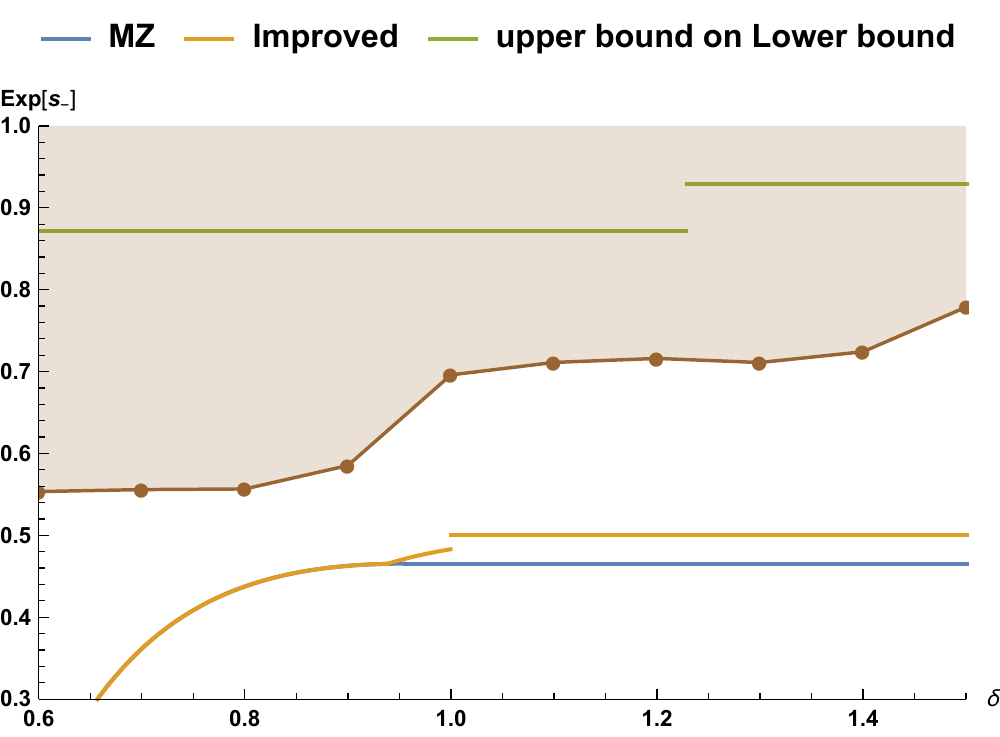}
\end{subfigure}
\caption{$\mathrm{Exp}[s_{\pm}]\ $ as a function of $\delta$, the half-width of the energy window. The blue line is the bound obtained in \cite{Baur}. The orange line denotes the improved bound that we report here. The green line is the analytical lower (upper) bound on the upper (lower) bound, while the brown dots stand for the lower (upper) bound on the upper (lower) bound obtained from enforcing the positive definiteness condition on the Fourier transform of $\pm(\phi_{\pm}-\Theta)$ via MATLAB. The bound on bounds represented by the green line is thus weaker than that of represented by the brown dots. The brown shaded region is \textbf{not} achievable by any bandlimited function.} 
\label{fig:mat}
\end{figure}

\begin{figure}[h!]
 \centering
\includegraphics[scale=0.4]{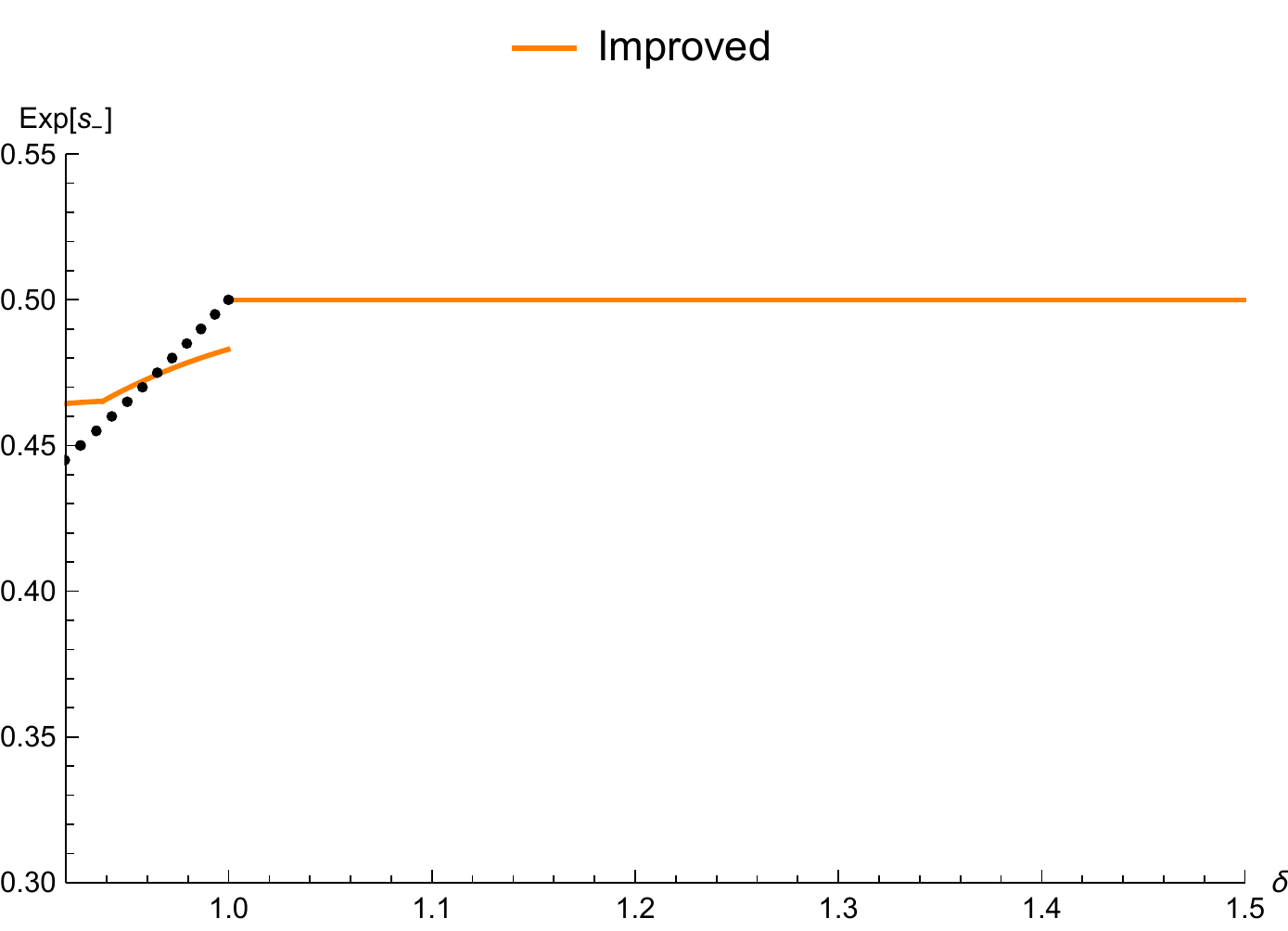}
\caption{$\mathrm{Exp}[s_{-}]\ $ as a function of $\delta$, the half-width of the energy window around $\delta=1$. The black curve makes the bounding curve continuous, i.e.\!~the bounding curve is now given by $\text{max}\left\{\text{orange curve},\text{black curve}\right\}$.} 
\label{fig:snew}
\end{figure}

Given this setup, improving the bound on order one correction in \eqref{eq:master} boils down to maximizing $\widehat{\phi}_{-}(0)$ and minimizing $\widehat{\phi}_{+}(0)$. Our results, regarding the bound on the $O(1)$ correction to the density of states are summarized by figure~[\ref{fig:mat}], where the green line and dots denote the lower (upper) bound on the upper (lower) bound. The orange lines denote the improved achievable bounds. The brown dots stand for the lower (upper) bound on the upper (lower) bound obtained from implementing the positive definiteness condition on the Fourier transform of $\pm(\phi_{\pm}-\Theta)$ via MATLAB. The bound on bounds represented by the green line is thus weaker than that of represented by the brown dots. In short, the brown shaded region is not achievable by any bandlimited function. In fact, if we use the analytic properties of the function involved on top of its bandlimited nature, then it is possible to show that the upper bound on the lower bound $c_-$ becomes $0.5/\delta$ i.e.\!~$\delta c_-\leq 0.5$ for $\delta\leq 1$.
One can notice a discontinuity in the graph of the lower bound around $\delta=1^{-}$. This indicates one can perhaps do better around that region. In fact, it turns out that one can do better in the interval $[1-\epsilon,1)$, for some positive $\epsilon$ (see figure.~\ref{fig:snew}). In the $\delta\to 1^{-}$, we analytically show that the lower bound is optimal. This optimality result follows from showing that the Monster CFT saturates this lower bound and thus independent of the fact whether we use bandlimited function or not.

\begin{figure}[h!]
 \centering
    \begin{subfigure}[b]{0.45\textwidth}
        \centering
\includegraphics[scale=0.7]{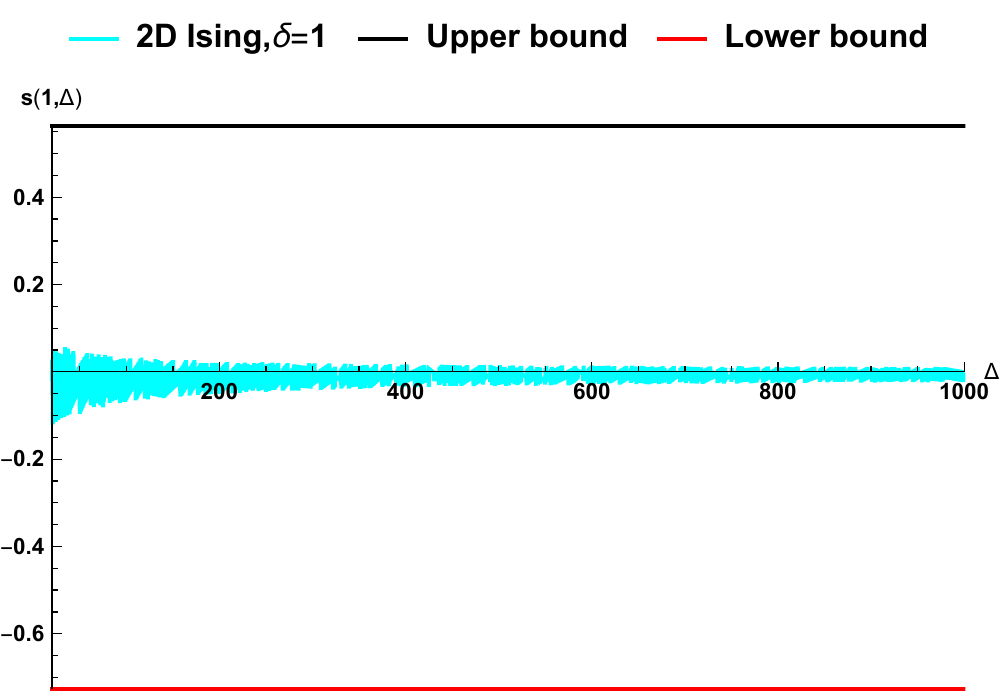}
\end{subfigure}
~
 \begin{subfigure}[b]{0.45\textwidth}
        \centering
\includegraphics[scale=0.7]{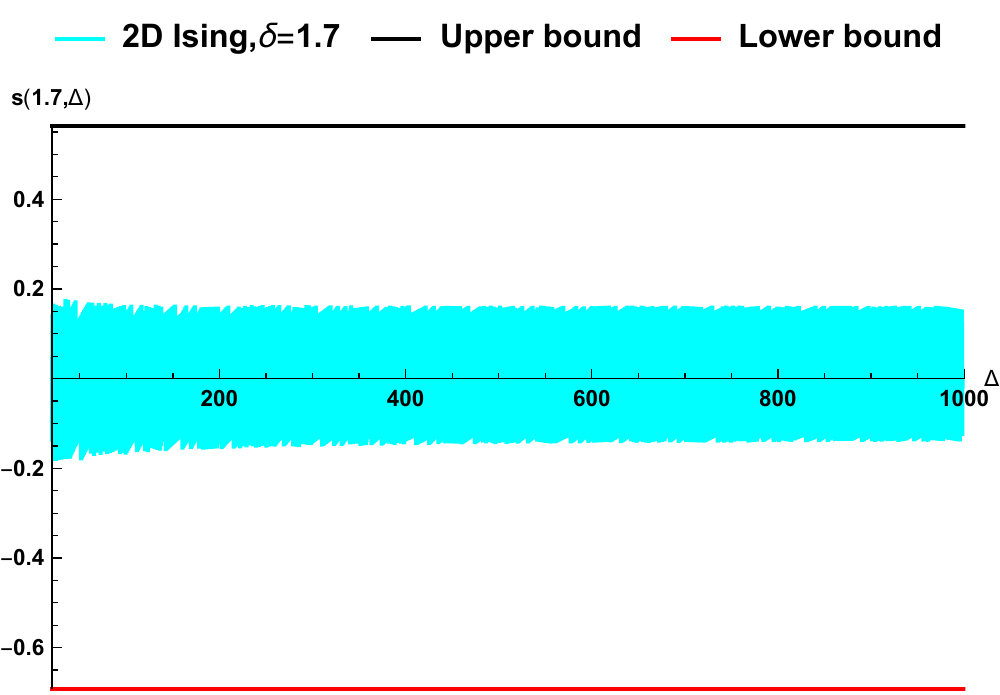}
\end{subfigure}
~
 \begin{subfigure}[b]{0.45\textwidth}
        \centering
\includegraphics[scale=0.7]{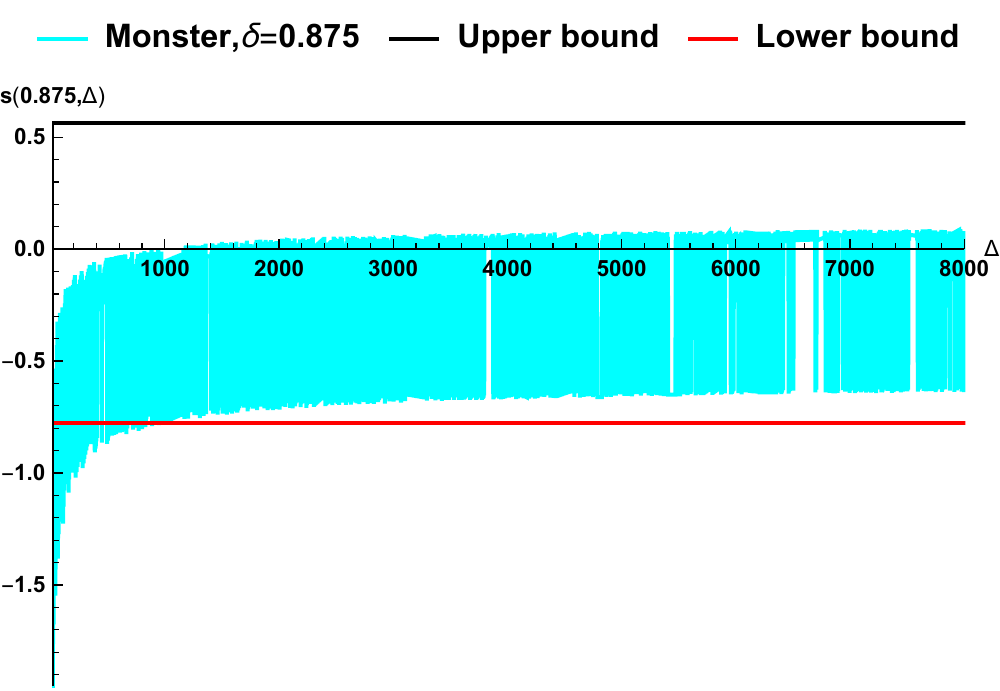}
\end{subfigure}
~
 \begin{subfigure}[b]{0.45\textwidth}
        \centering
\includegraphics[scale=0.7]{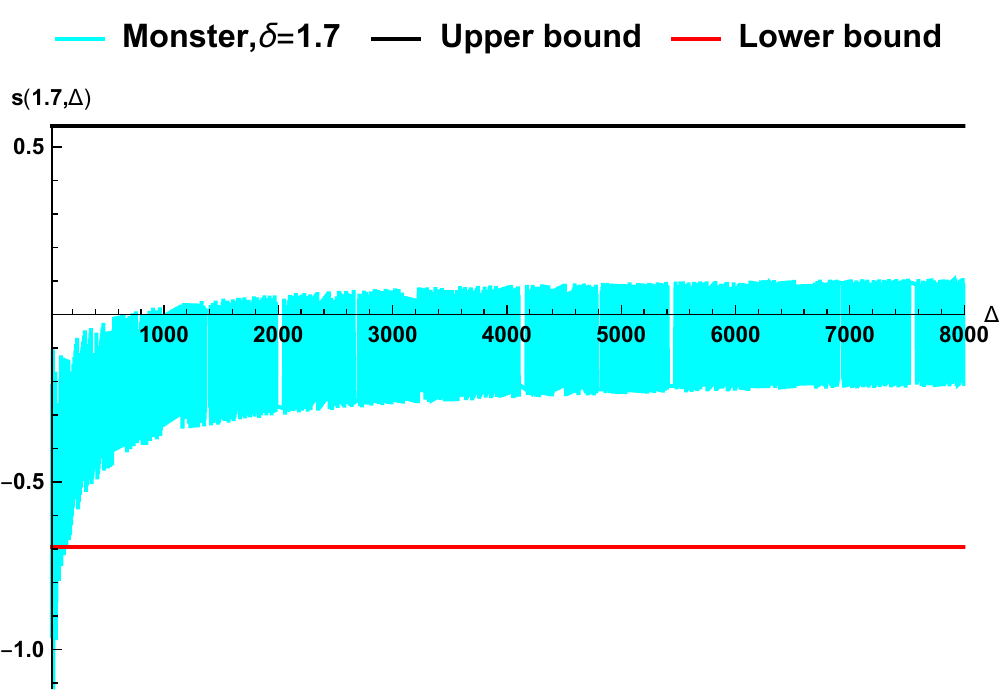}
\end{subfigure}
\caption{Verification of the bound on $s(\delta,\Delta)$, order one correction to entropy using $2$D Ising model (the top row) and non-chiral Monster CFT (the bottom row). We have plotted for different $\delta$, the half width of the interval under consideration. One can see for high enough $\Delta$, the bound is satisfied, indicating the asymptotic nature of the bound. The partition function for chiral Monster CFT can be reinterpreted as a S modular invariant particle function of a non chiral CFT with $c=12$. The dense cyan curve in between the red and the black line is obtained from the actual partition function.} 
\label{fig:verify}
\end{figure}

\begin{figure}[h!]
 \centering
    \begin{subfigure}[b]{0.5\textwidth}
        \centering
\includegraphics[scale=0.75]{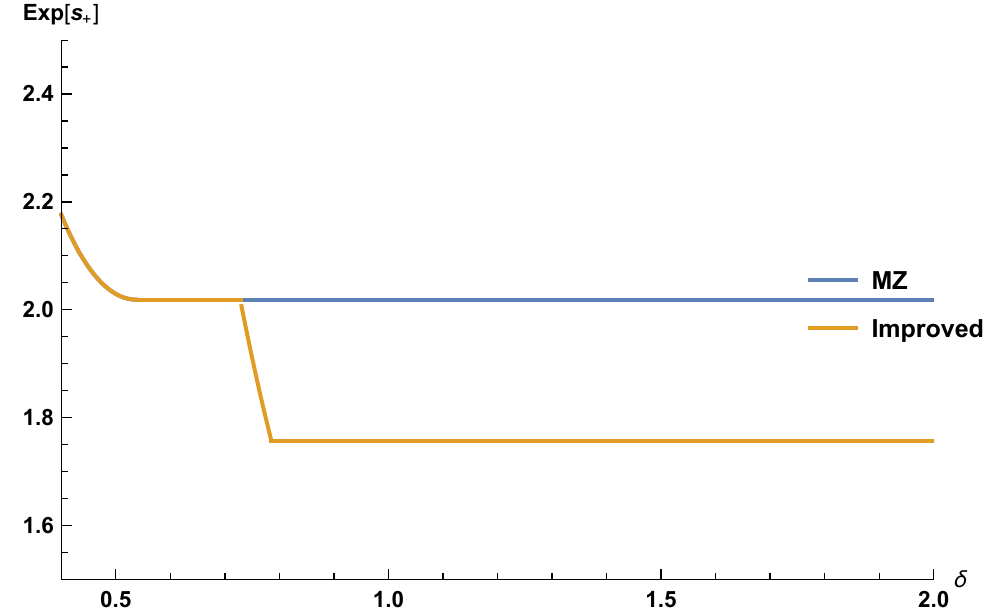}
  \end{subfigure}%
    ~ 
    \begin{subfigure}[b]{0.5\textwidth}
        \centering
\includegraphics[scale=0.75]{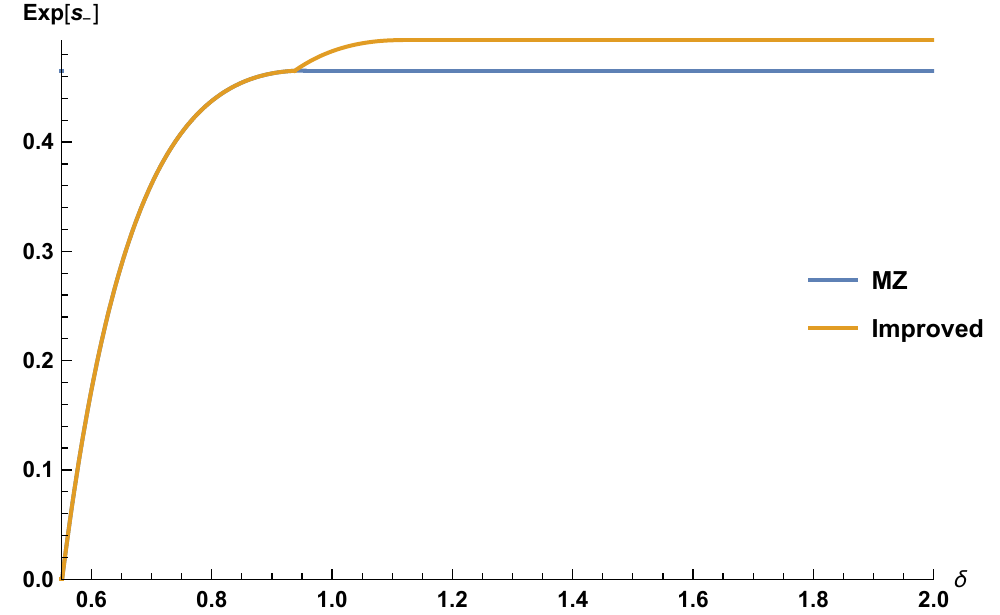}
 \end{subfigure}
\caption{$\mathrm{Exp}[s_{\pm}]\ $: The orange line denotes the improved lower (upper) bound while the blue line is from \cite{Baur}.} 
\label{fig:s_-}
\end{figure}

In particular, we show that the upper bound on $s(\delta,\Delta)$ is given by 
\begin{equation}\label{eq:785}
\exp\left[s_{+}(\delta)\right]=
\begin{cases}
MZ(\delta)\,,\ \ \delta< 0.73\\ \\
\frac{3}{40 \delta ^3} \left(11 \delta ^2+\frac{45}{\pi ^2}\right)\,,  \ \  0.73<\delta \le 0.785\\ \\
1.7578\,, \ \ \delta>0.785
\end{cases}
\end{equation}
where $MZ(\delta)$ is a function introduced in \cite{Baur} and defined as
\begin{align}
MZ(\delta)=\begin{cases}
\frac{\pi}{3}\left(\frac{\pi\delta}{2}\right)^3\left(\sin\left(\frac{\pi\delta}{2}\right)\right)^{-4}\,, 
\ \ \delta<\frac{a_{*}}{2\pi}\sim0.54
\\ \\
2.02\,,\ \ \delta>\frac{a_{*}}{2\pi}\sim0.54\,.
\end{cases}
\end{align}
Here, $a_*\sim 3.38$ satisfies $a_*=3\tan(a_*/4).$ We verify the new bound using the known partition function of $2$D Ising model and extremal CFTs as shown in figure~\ref{fig:verify}. Eq.\eqref{eq:785} is an improvement of the upper bound for $\delta>0.73$, as evident from figure~[\ref{fig:s_-}].\\

The lower bound $s_-(\delta)$ (except for some small interval $\delta\in [1-\epsilon,1)$) is given by
\begin{equation}\label{eq:784}
\exp\left[s_{-}(\delta)\right]=\begin{cases}
mz(\delta)\,,\quad  \frac{\sqrt{3}}{\pi}\le \delta < \frac{\sqrt{\frac{165}{19}}}{\pi }\sim 0.94, \\ \\
\frac{3 \left(11 \delta ^2-\frac{45}{\pi ^2}\right)}{40 \delta ^3}\,, \quad \frac{\sqrt{\frac{165}{19}}}{\pi } <\delta \le 1,\\ \\
0.5\,, \quad \delta>1.
\end{cases}
\end{equation}
where $mz(\delta)$ is a function, introduced in \cite{Baur} 
\begin{equation}
mz(\delta)=\begin{cases}
\frac{2 \left(\delta^2-\frac{3}{\pi^2}\right)}{3 \delta^3}\,,\quad \frac{\sqrt{3}}{\pi}\le \delta<\frac{3}{\pi}\sim 0.95,\\ \\
\frac{4\pi}{27}\sim 0.46\,, \quad \delta\ge \frac{3}{\pi}.
\end{cases}
\end{equation}
The eq.~\eqref{eq:784} is an improvement of the lower bound for $\delta>0.94$, as evident from figure~[\ref{fig:s_-}]. In that small interval, the better bound is depicted numerically in figure~[\ref{fig:snew}]. One can immediately verify the new bounds using the known partition function of $2$D Ising model and extremal CFTs as shown in figure~\ref{fig:verify}.\\

The rest of the paper details the derivation of the above. In section~\S\ref{der}, we derive the improvement on the bound on the $O(1)$ correction to the Cardy formula. The connection with sphere packing problem has been explored in section \S\ref{sec:spack}. The section~\S\ref{bonb} describes a systematic way to estimate how tight the bound can be made. We derive the optimal gap on the asymptotic spectra in section~\S\ref{optimal} and conclude with a brief discussion in section~\S\ref{conc}.

\section{Derivation of the improvement} \label{der}
The high energy density of states is controlled by the high temperature partition function. The usual trick of Cardy analysis involves writing down an expression for high temperature partition function and doing an inverse Laplace transformation as mentioned in the introduction. This provides us with a growth of the form $\left(\frac{c}{48\Delta^3}\right)^{1/4}\exp\left[2\pi\sqrt{\frac{c\Delta}{3}}\right]$. We shall call this Cardy-like growth. Nonetheless, this can not be the true answer as ``true'' density of states is a distribution. To obtain a rigorous version of the Cardy formula, we estimate the number of states appearing in order one window of width $2\delta$, centered at large energy $\Delta$ i.e.\!~we estimate
$$\int_{0}^{\infty}\text{d}\Delta'\ \rho(\Delta')\Theta\left(\Delta^\prime \in \left[\Delta-\delta,\Delta+\delta\right]\right)\,,$$
where $\Theta$ is the indicator function of the interval  $\left[\Delta-\delta,\Delta+\delta\right]$. We refer the readers to section $4$ of \cite{Baur} for details of the procedure leading to a bound when $\Delta$ goes to infinity. A nice and detailed exposition of the technique in context of asymptotics on $(h,\bar{h})$ plane can be found in \cite{Pal:2019zzr}. The notion of asymptote is far more rich on $(h,\bar{h})$ plane, hence the one can see the usefulness of Tauberian techniques in a detailed and transparent manner in such a scenario.  The key idea is to put a rigorous lower and upper bound on this quantity such that the bounds have similar Cardy-like growth as a function $\Delta$ and differ only by an order one multiplicative number. To do that, as mentioned briefly in the Introduction, we start with two bandlimited (functions of such kind have Fourier transform of bounded support) functions $\phi_\pm$ such that the following holds:
 \begin{align}
 \phi_{-}(\Delta^\prime) < \Theta\left(\Delta^\prime \in \left[\Delta-\delta,\Delta+\delta\right]\right) < \phi_+(\Delta^\prime)\,,
 \end{align}
 where $\Theta$ is the indicator function. Now, provided the Fourier transform of $\phi_{\pm}$ has a support on an interval which lies entirely within $\left[-2\pi,2\pi\right]$, one can derive in $\Delta\to\infty$ limit:
\begin{align}\label{basicone}
c_-\rho_0(\Delta) \le \frac{1}{2\delta}\int_{\Delta-\delta}^{\Delta+\delta}d\Delta^\prime\ \rho(\Delta^\prime) \le c_+ \rho_0(\Delta),
\end{align}
where $\rho_0(\Delta)$ reproduces the high temperature partition function i.e.\!~the contribution to the partition function from the vacuum in the dual channel. This is given by
\begin{align}
\rho_0(\Delta)= \pi\sqrt{\frac{c}{3}}\frac{I_{1}\left(2\pi\sqrt{\frac{c}{3}\left(\Delta-\frac{c}{12}\right)}\right)}{\sqrt{\Delta-\frac{c}{12}}}\Theta\left(\Delta-\frac{c}{12}\right)+\delta\left(\Delta-\frac{c}{12}\right).
\end{align}
We remark that one should only trust the leading piece of Bessel function. Thus the \eqref{basicone} is the same as \eqref{eq:master}. Furthermore, $c_{\pm}$ is defined as
\begin{align}\label{def:cpm}
c_{\pm}&=\frac{1}{2}\int_{-\infty}^{\infty}\ dx\ \phi_{\pm}(\Delta+\delta x).
\end{align} 
As mentioned earlier, $c_{\pm}$ are order one numbers. The condition that the Fourier transform of $\phi_{\pm}$ has a support on an interval which lies entirely within $\left[-2\pi,2\pi\right]$ ensures that one can ignore the contribution coming from heavy states toward the density of states when doing the high temperature expansion of the partition function, only the vacuum contribution survives in the eq.~\eqref{basicone}. With this constraint in mind, we consider the following functions:
\begin{align}
\phi_+(\Delta^\prime)&=\left[\frac{\sin \left(\frac{\Lambda_+ (\Delta^\prime -\Delta)}{6}\right)}{\frac{\Lambda_+(\Delta^\prime -\Delta)}{6}}\right]^6\left(1+\frac{(\Delta^\prime-\Delta)^2}{\delta^2}\right)\,,\\
\phi_-(\Delta^\prime)&=\left[\frac{\sin \left(\frac{\Lambda_-(\Delta^\prime -\Delta)}{6}\right)}{\frac{\Lambda_-(\Delta^\prime -\Delta)}{6}}\right]^6\left(1-\frac{(\Delta^\prime-\Delta)^2}{\delta^2}\right)\,.
\end{align}
In order to ensure that the indicator function of the interval $\left[\Delta-\delta,\Delta+\delta\right]$ is bounded above by $\phi_+$, we need to have 
\begin{align}\label{eq:cons1}
\delta\Lambda_+\le 4.9323\,.
\end{align}
The number in the eq.~\eqref{eq:cons1} is obtained by requiring that $\phi_{+}\left(\Delta\pm\delta\right)>1$. The functions $\phi_\pm$ have Fourier transforms with bounded supports $\left[-\Lambda_{\pm},\Lambda_{\pm}\right],$ respectively. Thus, in order for this support to lie within $[-2\pi, 2\pi],$ we also require that $\Lambda_{\pm}<2\pi$. The bound is then obtained by minimizing (or maximizing)
 \begin{align}
 c_\pm=\frac{1}{2\delta}\int d\text{x}\ \phi_\pm(\Delta+x)=\frac{3\pi \left(11\delta^2\Lambda_\pm^2\pm 180\right)}{20\delta^3\Lambda_\pm^3}
 \end{align}
 for a given $\delta$ by varying $\Lambda_{\pm}$ subject to the constraint given by the eq.~\eqref{eq:cons1}, as well as $\Lambda_\pm<2\pi$. From the eq.~\eqref{basicone}, one can conclude \cite{Baur} that
 \begin{align}
 c_- \le \exp\left[s(\delta,\Delta)\right] \le c_+.
 \end{align}
Since for a fixed $\delta$, $c_+$ is a monotonically decreasing function of $\Lambda_+$, we deduce that $c_+$ should be minimized by 
\begin{align}
\Lambda_+=\text{min}\left\{2\pi,\frac{4.9323}{\delta}\right\}=\begin{cases}
2\pi,\ \ \delta<0.785,\\
\frac{4.9323}{\delta},\ \ \delta>0.785.
\end{cases}
\end{align} 
This explains the number $0.785$ appearing in the bounds in the eq.~\eqref{eq:785}. The final bound can be obtained by combining these results with the result of \cite{Baur}. A similar analysis can be performed on $c_-$. These procedures yield the eq.~\eqref{eq:785} for the upper bound, while the lower bound is given by
\begin{equation}\label{eq:786}
\exp\left[s_{-}(\delta)\right]=\begin{cases}
mz(\delta)\,,\quad  \frac{\sqrt{3}}{\pi}\le \delta < \frac{\sqrt{\frac{165}{19}}}{\pi }\sim 0.94, \\ \\
\frac{3 \left(11 \delta ^2-\frac{45}{\pi ^2}\right)}{40 \delta ^3}\,, \quad \frac{\sqrt{\frac{165}{19}}}{\pi } <\delta < \frac{3\sqrt{\frac{15}{11}}}{\pi } \sim1.12, \\ \\
\frac{11}{60} \sqrt{\frac{11}{15}} \pi \sim 0.49\,, \quad \delta>\frac{3\sqrt{\frac{15}{11}}}{\pi } \sim 1.12.
\end{cases}
\end{equation}. 

\begin{figure}[h!]
 \centering
\includegraphics[scale=0.75]{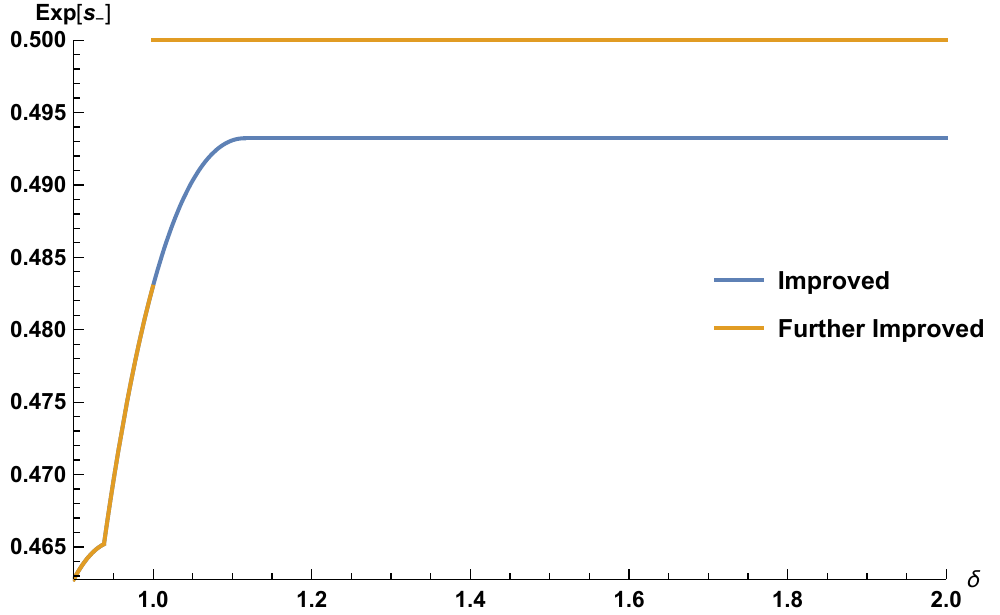}
\caption{The orange line represents the improvement on the lower bound by using the function $\phi^{\text{Sphere}}_-$ appearing in the sphere packing problem.}
\label{fig:spherepack}
  \end{figure}

The lower bound can be further improved for $\delta>1$ by considering the following function whose Fourier transform has a support over $[-\frac{2\pi}{\delta},\frac{2\pi}{\delta}].$
\begin{align}\label{magicfunc}
\phi^{\text{Sphere}}_-(\Delta^\prime) := \frac{1}{1-\left(\frac{\Delta^\prime-\Delta}{\delta}\right)^{2}}\left(\frac{\sin\left(\frac{\pi(\Delta^\prime-\Delta)}{\delta}\right)}{\frac{\pi(\Delta^\prime-\Delta)}{\delta}}\right)^{2}.
\end{align}
This yields $c_-=0.5$, which is an improvement over the above; see figure~\ref{fig:spherepack}.

 {\paragraph{Lower bound and its optimality in $\delta\to 1^{-}$:}
We can see that there is a discontinuity in the orange curve, as in figure~\ref{fig:spherepack}. It turns out that for $\delta<1$, an improvement is possible which makes the bounding curve continuous (see figure.~[\ref{fig:snew}]). The improvement is possible by considering the following function for $\sigma \in (0,1)$ and letting $\epsilon\to 1^{+}$ and $\sigma \to 1^{-}$ at at the end of the day: 
\begin{equation}\label{def:funcm}
\begin{aligned}
\phi_-(\Delta')=\frac{\sigma}{1-\left(\frac{\Delta^\prime-\Delta}{\epsilon}\right)^{2}}\left(\frac{\sin\left(\frac{\pi(\Delta^\prime-\Delta)}{\epsilon}\right)}{\frac{\pi(\Delta^\prime-\Delta)}{\epsilon}}\right)^{2}+\frac{(1-\sigma)\cos^{2}\left(\frac{\pi\left(\Delta^\prime-\Delta\right)}{\epsilon}\right)}{1-4\left(\frac{\Delta^\prime-\Delta}{\epsilon}\right)^2}
\end{aligned}
\end{equation} 
We consider this function in $\epsilon\to1^{+}$ limit, which enforces the constraint on the support of its Fourier transform i.e. we have $\Lambda \to (2\pi)^{-}$. Now, $f(x)\equiv \phi_-(\Delta+x)$ is negative for $|x|>\delta(\sigma)$, a function of $\sigma$. We can find out $\delta(\sigma)$ numerically from the function $f$ by noting that $\delta(\sigma)$ is the least positive root of the equation $f(x)=0$. This, in particular, shows that as $\sigma\to 1^{-}$, we have $\delta(\sigma)\to 1^{-}$, the new bound $c_-$ indeed approaches $0.5$.\\ 

Now let us show that $0.5$ is indeed the optimal value for $\delta \to1^{-}$. To achieve this, we will consider Monster CFT and a sequence of energy window centered at $\Delta_n=n$ and of half-width $\delta\to 1^{-}$. Clearly, the window contains all the states with $\Delta=n$ and nothing else. From the estimation of \cite{10.2307/1968796, 10.2307/2371313}, we know that the number of states with $\Delta=n$ is asymptotically given by $\rho_0(n)$. Thus we have
\begin{align}
\frac{1}{2\delta}\int_{n-\delta}^{n+\delta}\text{d}\Delta'\ \rho(\Delta') \simeq 0.5\rho_0(n)\,,\ \text{for}\ \delta\to 1^{-}
\end{align}
One can check this numerically as well using Monster CFT as depicted below in figure~[\ref{fig:verifysat}]. One needs to go to really high enough $\Delta$ to verify this.
\begin{figure}[h!]
 \centering
    \begin{subfigure}[b]{0.45\textwidth}
        \centering
\includegraphics[scale=0.5]{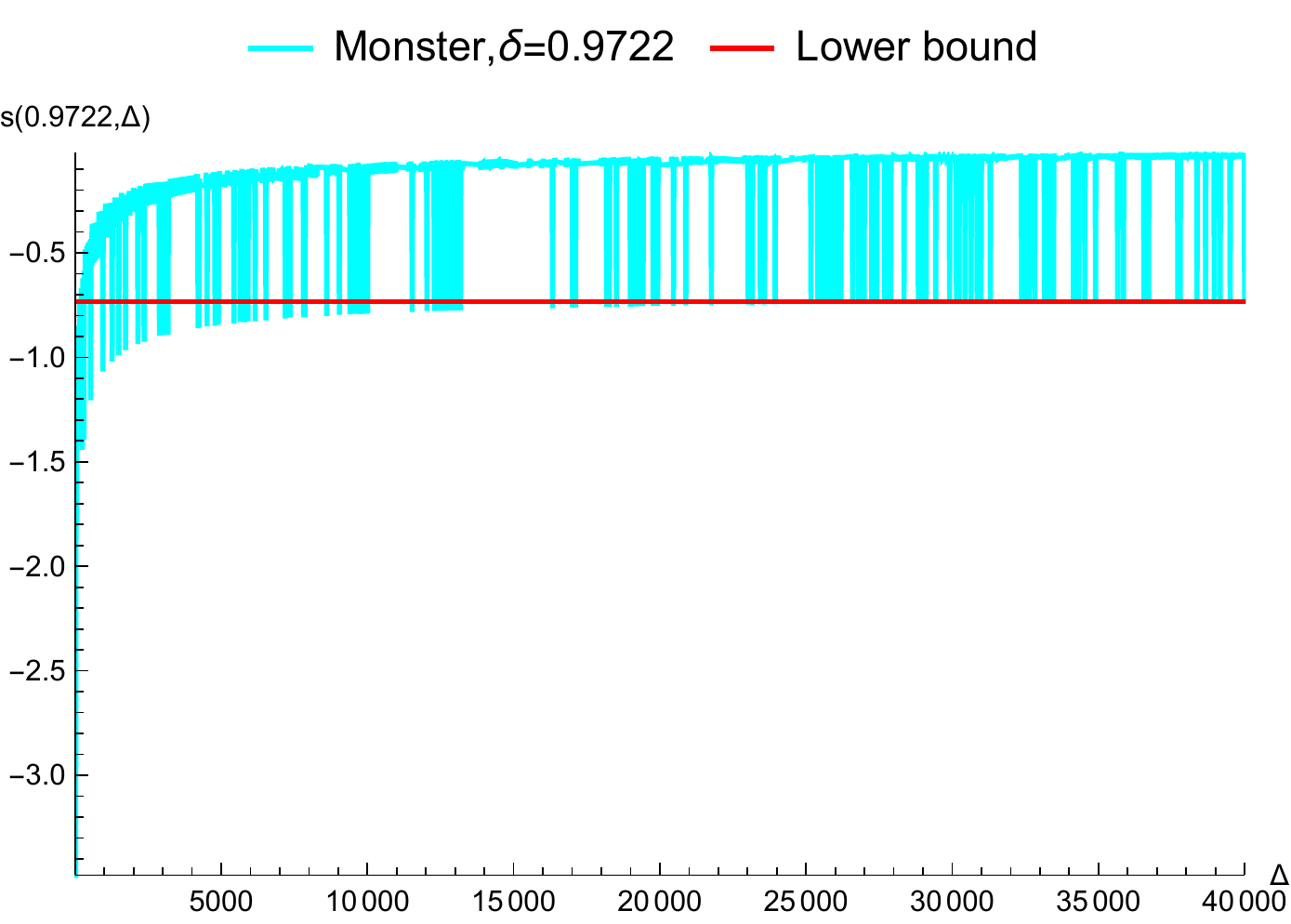}
\end{subfigure}
~
 \begin{subfigure}[b]{0.45\textwidth}
        \centering
\includegraphics[scale=0.5]{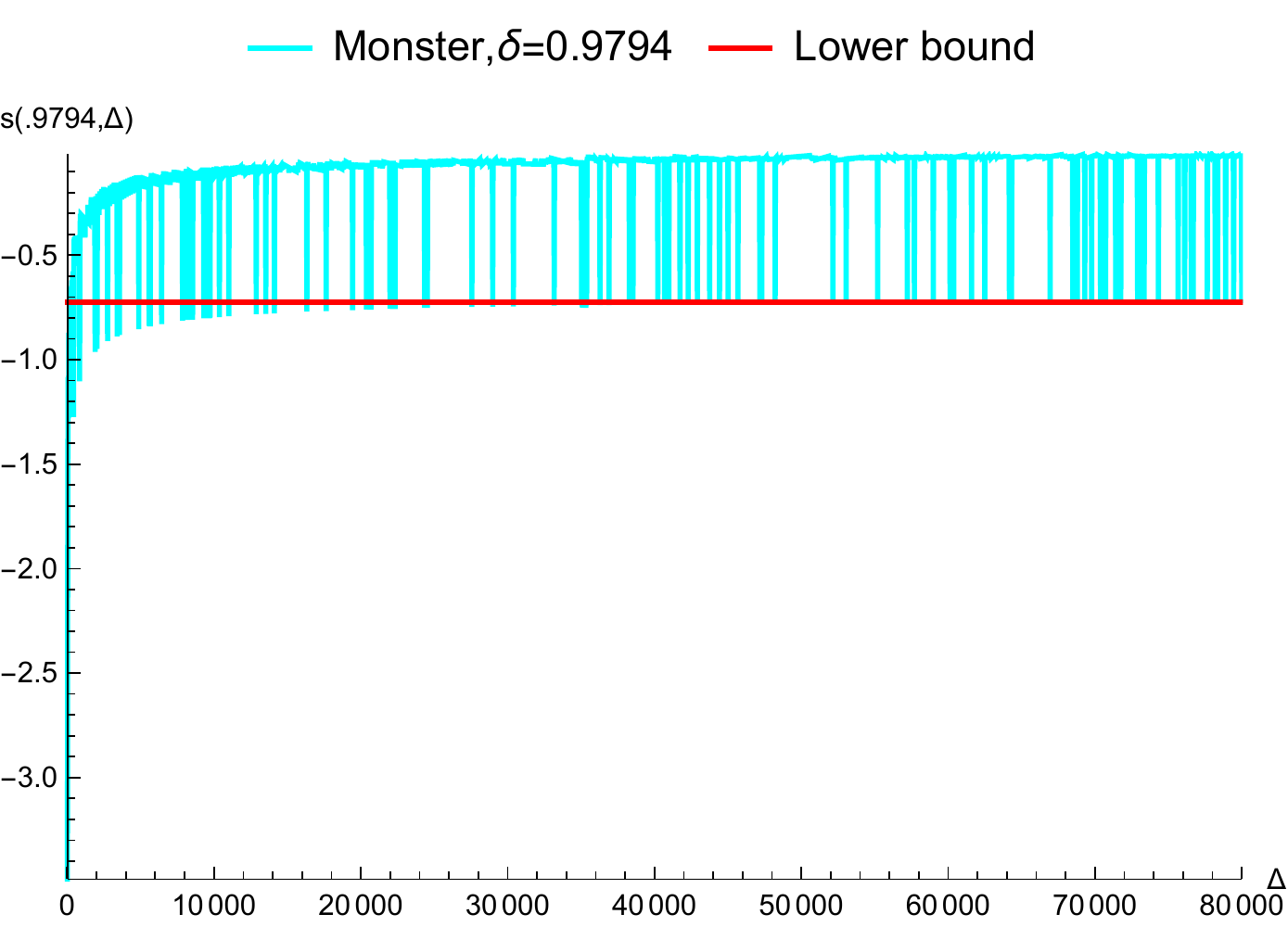}
\end{subfigure}
\caption{In cyan, we have $s(\delta,\Delta)$\  for Monster CFT as a function of $\Delta$, for $\delta=0.9722$ and $\delta=0.9794$. The red line denotes the lower bound using the function in eq.~\eqref{def:funcm}. The examples show the saturation of the bound, since the cyan line touches red bounding line for a sequence of $\Delta_k.$ } 
\label{fig:verifysat}
\end{figure}
Further verification can be obtained by looking at the subsequence $\Delta_n=n$ and verifying against the bound, as depicted in figure~[\ref{fig:verifysat2}].
\begin{figure}[h!]
 \centering
\includegraphics[scale=0.5]{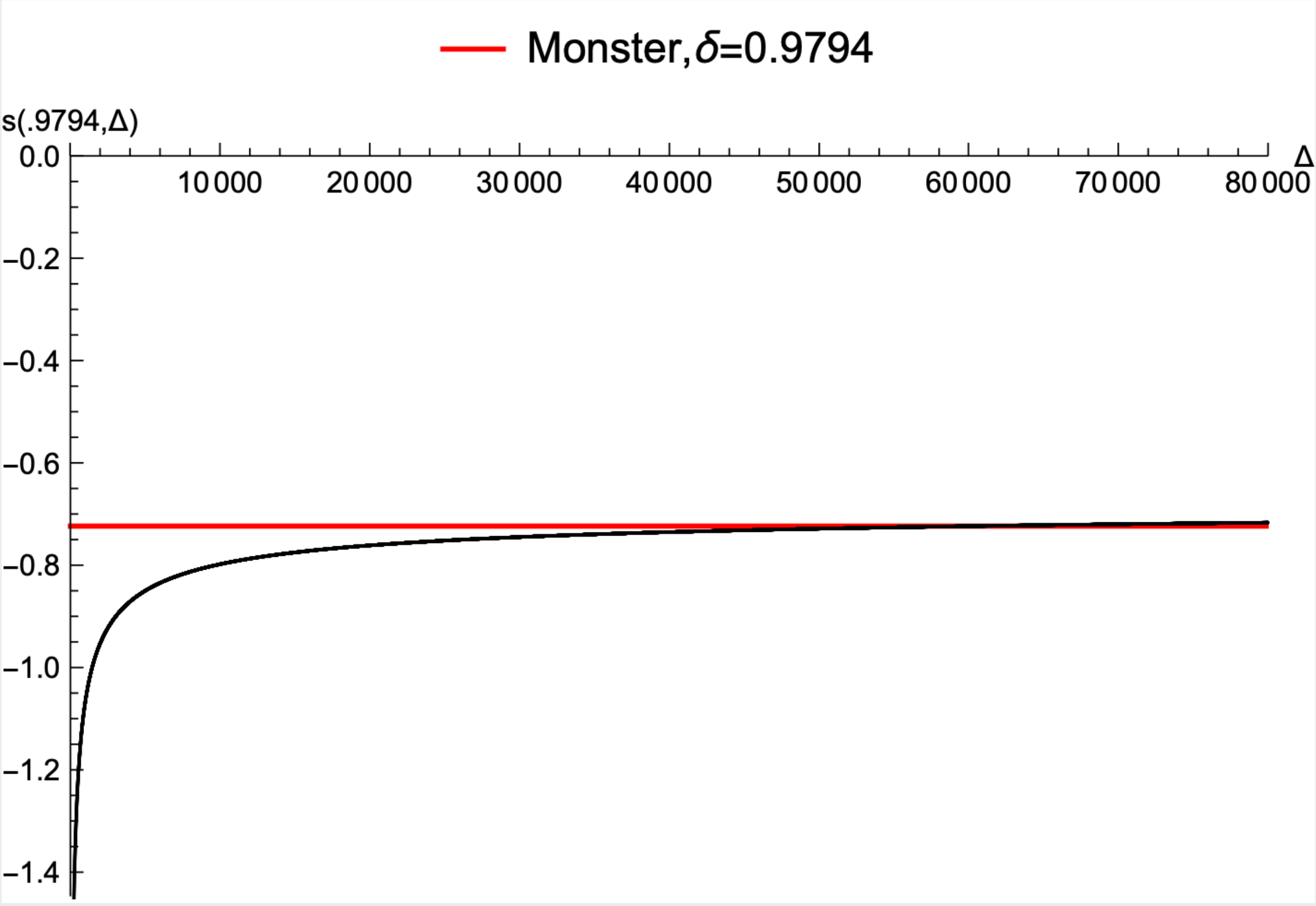}
\caption{In black, we have $s(\delta,\Delta_n=n)$\  for Monster CFT as a function of $n$ for $\delta=0.9794$. The red line denotes the lower bound using the function in eq.~\eqref{def:funcm}. The examples show the saturation of the bound, since the black line eventually touch the red bounding line. One can further see that the black line cross the red line which is a manifestation of the fact $\delta<1$. In fact it saturates to $\frac{0.5}{0.9794}$. We emphasize that the optimality is proven for $\delta\to1^{-}$ only. As we approach $\delta=1$ from below, the red line approaches $\log[0.5]$ and the black line asymptotes to $\log[0.5]$. } 
\label{fig:verifysat2}
\end{figure}
}

\section{Connection to the sphere packing problem}\label{sec:spack}
The purpose of this section is to elucidate a curious connection between the sphere packing problem and the problem of finding the lower bound $s_-(\delta)$. We start with the basic inequality 
$$2\pi\widehat{\phi}_{-}(0)\rho_0(\Delta) \leq \int_{\Delta-\delta}^{\Delta+\delta} \text{d}\Delta'\ \rho(\Delta')\,,$$
along with the definition $s_-(\delta)=\log c_-=\log[\frac{\pi}{\delta} \widehat{\phi}_{-}(0)]$. Furthermore, the function $\phi_-$ satisfies the following:
\begin{itemize}
\item The support of $\widehat{\phi}_{-}(t)$ is a subset of $[-2\pi,2\pi]$.
\item $\phi_-$ minorises\footnote{Ideally we have $\phi_-(\Delta')\leq \Theta(\Delta' \in [\Delta-\delta,\Delta+\delta])$. Nonetheless, we can always choose to work with $\phi_-(\Delta+x)$, which we write as $\phi_-(x)$ to reduce the amount of cluttering.} the indicator function of the interval $[-\delta,\delta]$ i.e. $\phi_-(x)\leq \Theta(x \in [\delta,\delta]$
\end{itemize}

$\bigstar$ Given this, our objective is to find out the maximum achievable value of $\widehat{\phi}_{-}(0)$ and show that it is achieved.

The maximization problem stated above has some uncanny similarity with the Sphere packing problem \footnote{S.P.\!~thanks John McGreevy for pointing to \cite{Hartman:2019pcd}, where sphere packing plays a pivotal role.} for $\delta\leq 1$. In the sphere packing problem, the analogue of $\phi_-$ and $\widehat{\phi}_{-}$ is played by the Fourier transform pair $f, \widehat{f}$, but they satisfy slightly different conditions: 
\begin{align}
f(x) &\le 0\ \text{for}\ |x|>1,\\
\widehat{f}(k)&\ge 0\,,\\
f(0)&=1\,.
\end{align}
In our case, we have $x\leftrightarrow \Delta^\prime-\Delta$ and $k\leftrightarrow t$. $\phi_-$ plays the role of $f$. The first condition is almost same as saying $\phi_-$ being the minoriser of the indicator function. The difference lies in the fact that the sphere packing problem does not restrict the function $f$ when $x\in [-1,1]$ whereas we want $\phi_-$ to be less than or equal to $1$ in this range. The second condition is entirely different from our case. In the sphere packing problem, non-negativity of $\widehat{f}$ is needed, whereas we want the analogous function, $\widehat{\phi}_-(t)$ to have bounded support. The third condition is a normalization choice in context of sphere packing whereas in our case we want $\phi_-(0)\leq 1$. Last but not the least, in both problems, the goal is to maximize $\widehat{f}(0)$ i.e.\!~$\widehat{\phi}_{-}(0)$. For more details on the relevance of sphere packing to CFT and exciting new results on modular bootstrap, we refer the readers to a recent paper \cite{Hartman:2019pcd}.\\
 
In one dimension \cite{cohn2003new}, where the sphere packing problem is trivial, the relevant optimal function $f$ is given by the eq.~\eqref{magicfunc}. This solves the sphere packing problem in one dimension since the Fourier transform is positive\footnote{For higher dimensions too, bandlimited functions are used (see, for example, Proposition 6.1 in \cite{cohn2003new}); nonetheless, they do not provide the tightest bound for the higher dimensional sphere packing problem. For $n=1$, the function appearing in the said proposition is related to the one that we have used. For other values of $n,$ we obtain bounds strictly less than $1/2$. We thank Tom Hartman for pointing this out.}. Futhermore, this function also has bounded support in the Fourier domain, thus it solves our problem as well. In the previous section, we have found that indeed for $\delta\to 1^{-}$, the optimal value of $2\pi\widehat{\phi}_{-}(0)=1$. For $\delta=1$, this corresponds to $c_-=0.5$ and $s_-=\log[0.5]$. This furnishes a concrete similarity between the solution of the problem appearing in the context of CFT and the solution of sphere packing problem in 1D via the magic function as given in eq.~\eqref{magicfunc}. We can further show that $2\pi\widehat{\phi}(0)\leq 1$ for $\delta\leq 1$. The key role in showing this will be played by the analytical property of the function $\phi_-$, coupled with theorems and techniques appearing in the sphere packing literature. In particular, we leverage the following theorem (it follows from modifying the corollary $3.2$ proven by Cohn in \cite{cohn2002new}: we remove the condition of non-negativity on $\widehat{f}$ and instead impose the bounded support condition, note that with this change, the function $f$ is no longer relevant for the sphere packing problem, nonetheless the theorem holds, which we state below):\\

\textit{$f$ is an admissible function i.e. there exists a $\kappa>0$ and $M>0$ such that
\begin{align}
|f(x)| \leq M (1+|x|)^{-1-\kappa}
\end{align}
Furthermore assume that $\widehat{f}(k)$ is in $L_1$ and has a bounded support, which is a subset of $[-2\pi,2\pi]$. Also assume that $f(x)\leq 0$ for $|x|\geq 1$. One can then show that\footnote{The relative factor of $2\pi$ comes because of the definition of Fourier transform. We define $f(x)=\int_{-\infty}^{\infty}\widehat{f}(k)e^{-\imath x k}$.}
\begin{align}
2\pi\widehat{f}(0) \leq f(0)\,.
\end{align}
}

Now we want to apply this theorem on $\phi_-$. We note that we want the function $\phi_-(x)$ appearing in the Tauberian analysis to be in $L_1$ and of exponential type to guarantee that $\widehat{\phi}_-$ has bounded support and finite everywhere inside its support. In particular, we would impose the technical assumption on $\phi_-$ that it is an admissible function. This automatically guarantees that $\phi_-$ is in $L_1$. Next, we note that $\widehat{\phi}_-$ is in $L_{\infty}$ (recall $\widehat{\phi}_-$ has bounded support and it is finite everywhere inside the support), in turn this implies that $\widehat{\phi}_-$ is in $L_p$ for all $p\in\mathbb{Z}_+$, in particular in $L_1$ because it has a bounded support. Now in CFT context, we also want $\phi_-$ to minorise the indicator function of the interval $[-\delta,\delta]$, in particular, this means that  $\phi_-(x)\leq 0$ for $|x|>\delta$.  Thus we see that $\phi_-$ belongs to the class of the functions considered in the theorem if $\delta\leq1$. In that scenario, we can apply the theorem and deduce 
\begin{equation}
\begin{aligned}
2\pi\widehat{\phi}_-(0)\leq \phi_-(0) \leq 1\,. 
\end{aligned}
\end{equation}
Thus the maximum possible value of $\pi\widehat{\phi}_{-}(0)$ is $0.5$ and we have shown that for $\delta\leq 1$, $2\pi\widehat{\phi}_-(0) \leq 1$ i.e. $c_-\leq \frac{1}{\delta}0.5$, and it is achievable for $\delta=1$.\\

\textit{Update: It turns out that the inequality $2\pi\widehat{\phi}_-(0) \leq \phi_-(0)$ can be obtained using Poisson summation formula without delving much into the details of technicalities of the theorem appearing in sphere packing paper. In fact, the key technique that goes into proving the theorem stated in the paper is the Poisson summation formula. Furthermore, the problem relevant for CFT is related to Beurling-Selberg problem. For $\delta=1$, the Beurling-Selberg problem and the Sphere packing problem admit the same optimizer and the bound coincides. For other values of $\delta$, one needs to leverage techniques appearing in the solution of Beurling-Selberg problem. For more details, we refer to \cite{Mukhametzhanov:2020swe}.}

 
 \section{Bound on bounds}\label{bonb} 
 In this section, we provide a systematic algorithm to estimate how tight the bounds can be made using bandlimited functions $\phi_\pm$. This provides us with a quantitative estimate\footnote{Ideally one would like to have a scenario, where the bound on bound becomes same as the achievable bound. This can indeed be done as shown by one of the authors in a different paper \cite{Mukhametzhanov:2020swe}. The numerics provided in this section is consistent (albeit less sharp) with the analytical result obtained in \cite{Mukhametzhanov:2020swe}.} of the limitation of the procedure which produces these bounds on the $O(1)$ correction to the Cardy formula. If one drops the requirement that the function be bandlimited, one might hope to do better. For the rest of this section, we will restrict ourselves to bandlimited functions only.\\
 
We recall that the functions $\phi_\pm$ are chosen in such a way that they satisfy
 \begin{align}
 \phi_{-}(\Delta^\prime) < \Theta\left(\Delta^\prime \in \left[\Delta-\delta,\Delta+\delta\right]\right) < \phi_+(\Delta^\prime)\,.
 \end{align} 
 This inequality gives a trivial bound on $c_\pm$:
 \begin{align}\label{ineq}
 c_-\le 1 \le c_+\,.
 \end{align}
 In what follows, we make this inequality tighter. In this context, the following characterization of the Fourier transform of a positive function in terms of a \textit{positive definite} function turns out to be extremely useful. Before delving into the proof, let us define the notion of positive definiteness of a function. Unless otherwise specified, here we will be dealing with functions from the real line to the complex plane. A function $f(t)$ is said to be positive definite if for every positive integer $n$ and for every set of distinct points $t_1, \ldots, t_n$ chosen from the real line, the $n\times n$ matrix $A$ defined by 
 \begin{align}
 A_{ij}= f(t_i-t_j)
 \end{align}
 is positive definite. A function $g(\Delta)$ is said to be positive if $g(\Delta)>0$ for every $\Delta$. One can show that the Fourier transform of a positive function is positive definite\footnote{The proof is given in a box separately at the end of this subsection for those who are interested.}. Now, let us explore how this characterization can improve the eq.~\eqref{ineq}. Without loss of generality, we set $\Delta=0$ henceforth, and define 
\begin{align}
g_{\pm}(\Delta^\prime)= \pm\left[\phi_\pm(\Delta^\prime)-\Theta\left(\Delta^\prime \in \left[-\delta,\delta\right]\right)\right].
\end{align}
At this point we use the fact that $\phi_{\pm}$ is a bandlimited function, i.e., it has a bounded support $\left[-\Lambda_\pm,\Lambda_\pm\right],$ and that $\Lambda_\pm < 2\pi$. This requirement stems from the procedure followed in \cite{Baur}. Thus we arrive at the following:
\begin{align}
\widetilde{g}_{\pm}(0)&= \pm2\delta \left(c_{\pm} -1\right),\\
\widetilde{g}_{\pm}(t)&= \mp 2\delta\left(\frac{\sin(t\delta)}{t\delta}\right)\, \text{for}\ |t| \ge 2\pi. 
\end{align}
The eq.~\eqref{ineq} states that $\tilde{g}(0)/2\delta>0$. In order to improve this, we construct $2\times 2$ matrices with $t_2 > 2\pi$:
\begin{align}
G^{(2)}_{\pm}=\begin{bmatrix}
\widetilde{g}_{\pm}(0) & \widetilde{g}_{\pm}(t_2)\\
\widetilde{g}_{\pm}(t_2) & \widetilde{g}_{\pm}(0)
\end{bmatrix}\,.
\end{align}
For a fixed $\delta$, we consider the first positive peak of $\tilde{g}_{\pm}$ outside $t>2\pi$. If this occurs at $t=t(\delta) $, we choose $t_2=t(\delta)$. Subsequently, the positive definiteness of the matrix  $G^{(2)}_{\pm}$ boils down to the inequality 
\begin{align}
\widetilde{g}_{\pm}(0)> \widetilde{g}_{\pm}\left(t(\delta)\right),
\end{align}
where $t(\delta)$ is the first positive peak of $\widetilde{g}_{\pm}$ outside $t>2\pi$. For example, we can show that (see the green lines in Fig.~\ref{fig:boundonbound}):
\begin{figure}[h!]
\centering
 \begin{subfigure}[b]{0.45\textwidth}
        \centering
\includegraphics[scale=0.7]{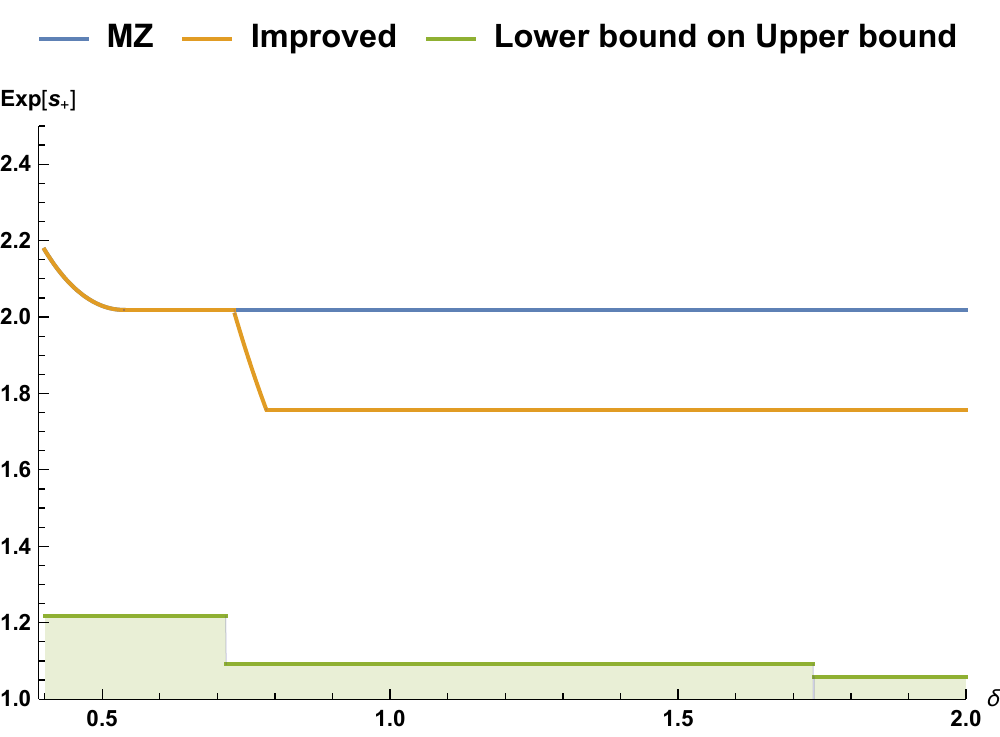}
\end{subfigure}
~~~~
 \begin{subfigure}[b]{0.45\textwidth}
        \centering
\includegraphics[scale=0.7]{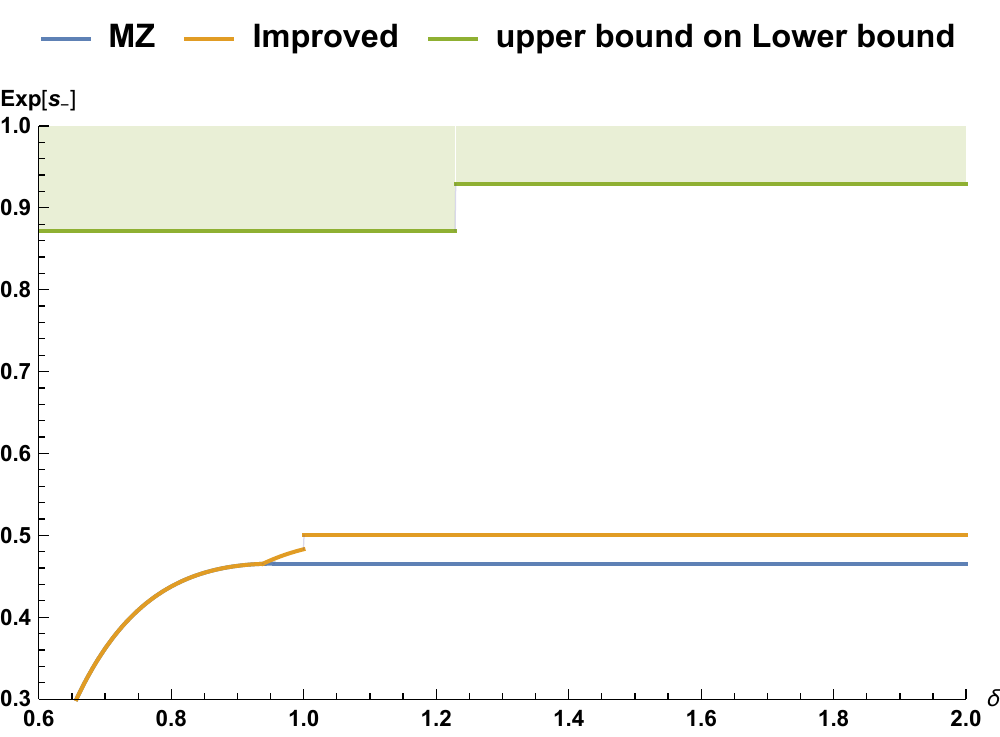}
\end{subfigure}
\caption{$\mathrm{Exp}[s_{\pm}]\ $: The green line is the analytical  lower and upper bound on upper and lower bound i.e. $c_{\pm}$ respectively. The green shaded region is not achievable by any bandlimited function.} 
\label{fig:boundonbound}
\end{figure}

\begin{align}
c_+ &> \begin{cases} 1.2172\,,&\ \delta < 0.715, \\
1.0913\,,&\ 1.735>\delta > 0.715,\\
1.0579\,,&\ 2.74>\delta > 1.736,\\
\end{cases}\\
 c_-&< \begin{cases}
 0.872\,,&\ \delta < 1.229, \\
  0.9291\,,&\ 2.238>\delta > 1.229. 
 \end{cases}
\end{align}

 \begin{ex}[Positive function $\Leftrightarrow$ Positive definite Function]{Fourier transform}{}
We will show that the Fourier transform of an even and positive function is a positive definite function. Consider a function $g(\Delta)$ and let us define the Fourier transform as
\begin{align}
\widetilde{g}(t)=\int_{-\infty}^{\infty} dt\ g(\Delta)e^{-\imath\Delta t} = 2\int_{0}^{\infty}dt\ \cos(\Delta t) g(\Delta).
\end{align}
Now, we construct the matrix
\begin{align}
G_{ij}= g(t_i-t_j)=2\int_{0}^{\infty}dt\ \cos\left[\Delta (t_i-t_j)\right] g(\Delta).
\end{align}
In order to show that $G$ is a positive definite matrix, i.e., $\sum_{ij}v_iv_jG_{ij}>0$ for $v_i \in \mathbb{R}$ such that $\sum_i v_i^2\neq 0$, we think of an auxiliary $2$ dimensional space with $n$ vectors $\vec{v}_{(i)},$ (for clarity, we remark that $i$ labels the vector itself, not its component) such that we have 
\begin{align}
\vec{v}_{(i)} \equiv \left(|v_i|\cos(\Delta t_i),|v_i|\sin(\Delta t_i)\right).
\end{align}
Thus, we have
\begin{align}
\sum_{ij}v_iv_jG_{ij}&=2\int_{0}^{\infty}dt\ \left(\sum_{ij}v_iv_j\cos\left[\Delta (t_i-t_j)\right] \right) g(\Delta)\\
&=2\int_{0}^{\infty}dt\ \left(\vec{V}\cdot\vec{V}\right)\ g(\Delta) >0 
\end{align}
if $t_1, \ldots, t_n$ are distinct. Here, $\vec{V}$ is given by 
\begin{align}
\vec{V}= \sum_{i} \mathrm{sign}(v_i)\ \vec{v}_{(i)}\,.
\end{align}
This completes the proof that the Fourier transform of an even positive function is a positive definite function. First of all, it is easy to see that $c_\pm,$ and hence the inequality, is insensitive to the midpoint of the interval, i.e., $\Delta$, so we set it to $0$ and this makes the functions $\phi_\pm$ and $\Theta$ even. In particular, we will be applying this theorem to $\phi_+(\Delta^\prime) -\Theta\left(\Delta^\prime \in \left[\Delta-\delta,\Delta+\delta\right]\right)$ and $\Theta\left(\Delta^\prime \in \left[\Delta-\delta,\Delta+\delta\right]\right)-\phi_-(\Delta^\prime)$. We make one more remark before exploring the consequences of this. The above result is true for any function, not necessarily even. The converse is also true due to \emph{Bochner's Theorem}, but in what follows, we do not require the converse statement.
 \end{ex}

\subsection*{Matlab implementation}
We implement the above argument using more than two points and making sure that $|t_i-t_j| \ge 2\pi$. For a fixed $\delta$, we use a random number generator to sample the points $t_i$ with the mentioned constraint. We do this multiple times and each time, we test the positive definiteness of the matrix $G$ by providing as an input the value of $\pm(c_{\pm}-1)$. The range of $\pm(c_{\pm}-1)$ is chosen to be from the first peak $t(\delta)$ till some value larger than the achievable bound given in \eqref{eq:785} and \eqref{eq:786}. This in turn yields a lower bound (or upper bound) for $c_\pm$ for each trial\footnote{We assume that the mesh size for $c_\pm-1$ is small enough that one can safely find out a lower bound.}. Subsequently, we pick out the best possible bound among all the trials. For example, we provide a table~[\ref{table:1}] showing the outputs from a typical run for improving the bound on the upper bound. The tables~[\ref{table:1}] and [\ref{table:2}]  improve the lower (upper) bound for $c_\pm$ and this is shown in the figure~[\ref{fig:mat}], where the brown dots are the stronger bounds over the green lines and disallow a larger region.

\begin{table}[h!]
\centering
\begin{tabular}{|c|c|c|c|c|}
\cline{1-5}
$\delta$ & Number of iterations & \# points & $\text{Max}(c_+)$ & Lower Bound \\ \cline{1-5}
0.4 & 10000 & 300 & 2.2 & 1.7042\\ \cline{1-5}
0.5 & 1000 & 300 & 2.02 & 1.6905 \\ \cline{1-5}
0.5 & 10000 & 200 & 2.02 & 1.7002 \\ \cline{1-5}
0.5 & 10000 & 300 & 2.02 & 1.7179 \\ \cline{1-5}
0.6 & 1000 & 200 & 2.02 & 1.6086 \\ \cline{1-5}
0.6 & 10000 & 200 & 2.02 & 1.5917 \\ \cline{1-5}
0.7 & 10000 & 200 & 2.02 & 1.4246 \\ \cline{1-5}
0.7 & 10000 & 250 & 2.02 & 1.4270 \\ \cline{1-5}
0.8 & 10000 & 200 & 1.757 & 1.3692 \\ \cline{1-5}
0.8 & 10000 & 200 & 2.757 & 1.3698 \\ \cline{1-5}
0.9 & 10000 & 200 & 2.757 & 1.3798 \\ \cline{1-5} 
1 & 20000 & 200 & 1.757 & 1.3759 \\ \cline{1-5}
1.1 & 10000 & 200 & 2.757 & 1.3331\\ \cline{1-5}
1.20 & 10000 & 150 & 2.757 & 1.2597\\
\cline{1-5}
1.25 & 10000 & 150 & 2.757 & 1.2581\\
\cline{1-5}
1.3 & 10000 & 170 & 2.757 & 1.2531\\
\cline{1-5}
1.4 & 10000 & 150 & 2.757 & 1.2581\\
\cline{1-5}
1.5 & 10000 & 150 & 1.757 & 1.2599\\ \cline{1-5}
1.5 & 10000 & 150 & 2.757 & 1.2597\\
\cline{1-5}
1.5 & 10000 & 150 & 2.757 & 1.2597\\
\cline{1-5}
1.6 & 10000 & 150 & 1.757 & 1.2313\\
\cline{1-5}
1.7 & 10000 & 150 & 1.757 & 1.1933\\
\cline{1-5}
\end{tabular}
\caption{Typical output from a run yielding lower bounds for the upper bound $c_+$. The $\mathrm{Max}(c_+)$ column contains a number that is greater than or equal to what can already be achieved.}
\label{table:1}
\end{table}

\begin{table}[h!]
\centering
\begin{tabular}{|c|c|c|c|c|}
\cline{1-5}
$\delta$ &Iteration Number & \# points & $\text{Min}(c_-)$ & Upper Bound \\ \cline{1-5}
0.6 & 1000 & 200 & 0.173  &  0.5738\\ \cline{1-5}
0.6 & 10000 & 200 & 0.173 & 0.5535  \\ \cline{1-5}
0.7 & 10000 & 200 & 0.362 &  0.5604\\ \cline{1-5}
0.7 & 10000 & 250 & 0.362 & 0.5559 \\ \cline{1-5}
0.8 & 10000 & 200 & 0.44 & 0.5567\\ \cline{1-5}
0.9 & 10000 & 200 & 0.46 & 0.5853 \\ \cline{1-5} 
1 & 10000 & 200 & 0.48 & 0.6960 \\ \cline{1-5}
1.1 & 10000 & 200 & 0.49 & 0.7112\\ \cline{1-5}
1.2 & 10000 & 150 & 0.49 & 0.7161\\
\cline{1-5}
1.2 & 10000 & 180 & 0.49 & 0.7161\\
\cline{1-5}
1.3 & 10000 & 170 &  0.49& 0.7111 \\
\cline{1-5}
1.4 & 10000 & 150 &  0.49 & 0.7243\\
\cline{1-5}
1.5 & 10000 & 150 &  0.49 & 0.7788 \\ \cline{1-5}
\cline{1-5}
1.6 & 20000 & 150 &  0.49 & 0.7895 \\ \cline{1-5}
\cline{1-5}
1.7 & 20000 & 150 &  0.49 & 0.7861 \\ \cline{1-5}
\cline{1-5}
\end{tabular}
\caption{Typical output from a run providing upper bound for the lower bound $c_-$. The $\text{Min}(c_-)$ column contains a number that is smaller than or equal to what can already be achieved.}
\label{table:2}
\end{table}


\section{Bound on spectral gap: the Optimal one}\label{optimal}
In this section, we switch gear and explore the asymptotic spectral gap. In \cite{Baur}, it has recently been shown that the asymptotic gap between Virasoro primaries are bounded above by $2\sqrt{\frac{3}{\pi^2}}\simeq 1.1$ and it has been conjectured that the optimal gap should be $1$. The example of Monster CFT tells us that the gap cannot be below than $1$, hence $1$ should be the optimal number. In this section, we show that the previous bound $2\sqrt{\frac{3}{\pi^2}}$ can be improved and made arbitrarily closer to the optimal value $1$. The idea of the gap comes from binning the states and putting a positive lower bound on that. If the width of the bin is very small, one finds that the lower bound on that number of states becomes negative. This indicates that if the bin width is very small, we might land up with no states in the bin. Thus we need to find a minimum bin width which would still allow us to prove a positive lower bound. Ideally, to prove this one should find out a function $f$ (which will eventually play the role of $\phi_-$ in this game, to be precise $f(\Delta^\prime)=\phi_-(\Delta+\Delta^\prime)$) such that the following holds:
\begin{align}
f(\Delta^\prime) &\le \Theta\left(\Delta^\prime \in \left[-\frac{\epsilon}{2},\frac{\epsilon}{2}\right]\right)
\end{align}
and
\begin{align}
\tilde{f}(t)&=0\ \text{for}\ |t|\ge\frac{2\pi}{\epsilon}\,,\ \epsilon>1\\
\tilde{f}(0)&>0
\end{align}
This would have implied 
\begin{align}
\int_{\Delta-\delta}^{\Delta+\delta}d\Delta^\prime \rho(\Delta^\prime) > 0
\end{align}
Now what would happen if $\tilde{f}(0)=0\ $? One needs to go back to the original derivation and reconsider it carefully. Hence instead of the eq.~\eqref{basicone}, we consider a more basic inequality\cite{Baur}:
\begin{align}\label{inequality}
\nonumber &\exp\left[\beta(\Delta-\delta)\right]\int d\Delta^\prime \rho_0(\Delta^\prime)e^{-\beta\Delta^\prime}\phi_-(\Delta^\prime) - Z_{H}\left(\frac{4\pi^2\beta}{\beta^2+\Lambda_-^{2}}\right)e^{-\beta \frac{c}{12}}\int_{-\Lambda_-}^{\Lambda_-}dt\ |\widehat{\phi}(t)|\\
&\le \int_{\Delta-\delta}^{\Delta+\delta}d\Delta^\prime \rho(\Delta^\prime)
\end{align}
where $\Lambda_{-}=\frac{2\pi}{\epsilon}$ and $Z_{H}(\beta)$ is the contribution from the heavy states and defined as
\begin{align}
Z_{H}(\beta)=\sum_{\Delta>\Delta_H>\frac{c}{12}}e^{-\beta\left(\Delta-\frac{c}{12}\right)}\,.
\end{align}
 Now we make the following choice for $\phi_-$:
\begin{align}
\phi_-(\Delta^\prime)=\frac{\cos^{2}\left(\frac{\pi\left(\Delta^\prime-\Delta\right)}{\epsilon}\right)}{1-4\left(\frac{\Delta^\prime-\Delta}{\epsilon}\right)^2}\,, \quad f(\Delta^{\prime})=\frac{\cos^{2}\left(\frac{\pi\Delta^\prime}{\epsilon}\right)}{1-4\left(\frac{\Delta^\prime}{\epsilon}\right)^2}\
\end{align}
This function $f$ has the following properties: 
\begin{align}
f(\Delta^\prime)&\le \Theta\left(\Delta^\prime \in \left[-\frac{\epsilon}{2},\frac{\epsilon}{2}\right]\right)\\
\tilde{f}(t)&=0\quad \text{for}\quad |t|\ge \frac{2\pi}{\epsilon}\\
\tilde{f}(0)&=0\quad \Rightarrow\quad c_-=0
\end{align}
Since $c_-=0$, one cannot readily evaluate the integral appearing in \eqref{inequality} by saddle point method and deduce $\exp\left[\beta(\Delta-\delta)\right]\int d\Delta^\prime \rho_0(\Delta^\prime)e^{-\beta\Delta^\prime}\phi_-(\Delta^\prime) = c_-\rho_0(\Delta)$, so we look for subleading corrections to the saddle point approximation. We find that the leading behavior is given by, after setting $\beta=\pi \sqrt{\frac{c}{3\Delta}}$, 
\begin{align}
\exp\left[\beta(\Delta-\delta)\right]\int d\Delta^\prime \rho_0(\Delta^\prime)e^{-\beta\Delta^\prime}\phi_-(\Delta^\prime) =C\rho_0(\Delta)\,,
\end{align}
where $C$ turns out to be
\begin{align}
C=\int_{0}^{\infty} dx\  \left(\frac{\cos^{2}\left(\pi \frac{x}{\epsilon}\right)}{1-4\frac{x^2}{\epsilon^2}}\right) \exp\left[\frac{-x^2}{2\pi\sqrt{\frac{c}{3}}\Delta^{\frac{3}{2}} }\right]\,.
\end{align}
We remark that $C>0$ for any finite $\Delta$ and it becomes $0$ only at infinitely large $\Delta$. The second piece in the eq.~\eqref{inequality} for large $\Delta$ goes as $\rho_0(\Delta)^{1-\frac{1}{2}\left(1-\frac{1}{\epsilon^2}\right)}$. The analysis for this second term is exactly the same as done in \cite{Baur}. For sufficiently large $\Delta$, it can be numerically verified that $\rho_0(\Delta)^{1-\frac{1}{2}\left(1-\frac{1}{\epsilon^2}\right)}$ is subleading compared to $C\rho_0(\Delta)$ as long as $\epsilon>1$ (we also provide an analytical proof later on). Here we have 
\begin{align}
\rho_0(\Delta) \underset{\Delta\to\infty}{=} \left(\frac{c}{48\Delta^3}\right)^{\frac{1}{4}}\exp\left[2\pi\sqrt{\frac{c\Delta}{3}}\right]\,.
\end{align}

\begin{figure}[h!]
\centering
\includegraphics[scale=0.85]{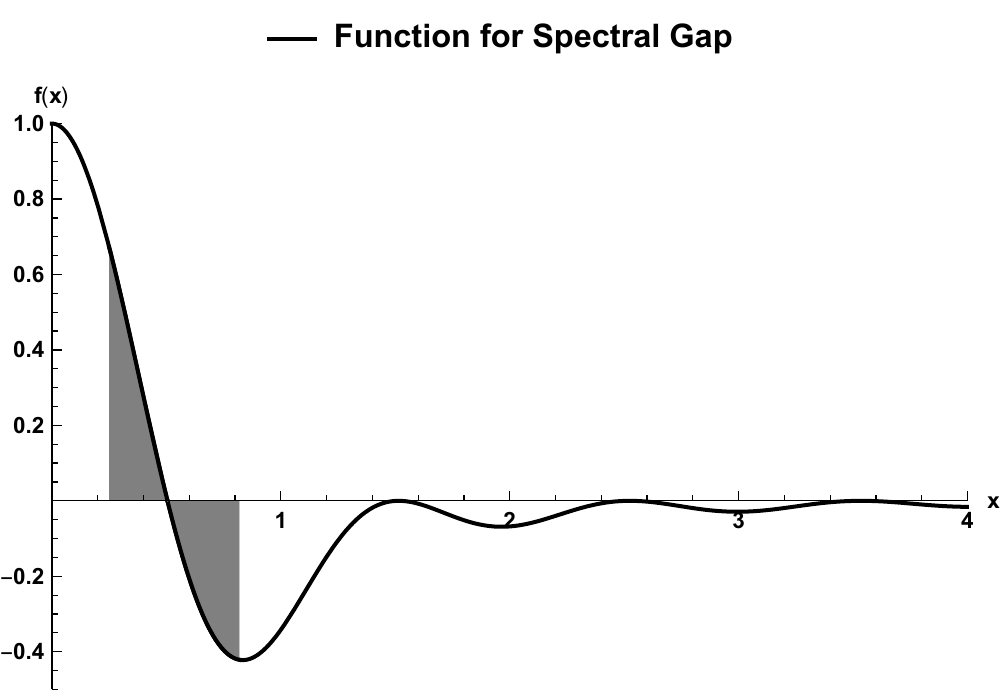}
\caption{ The function $\left(\frac{\cos^{2}\left(\pi \frac{x}{\epsilon}\right)}{1-4\frac{x^2}{\epsilon^2}}\right)$, the shaded region is the area under the function restricted to the interval $[\epsilon_1,\epsilon_2]$. Here $\epsilon_1=0.25,\epsilon=1.01,\epsilon_2=0.819$. These are chosen to ensure the shaded area is $0$.} 
\label{fig:func}
\end{figure}

One can analytically show that $\rho_0(\Delta)^{1-\frac{1}{2}\left(1-\frac{1}{\epsilon^2}\right)}$ is subleading to $C\rho_{0}(\Delta)$ for large $\Delta$. One way to show this is to have an estimate for $C$. We start with the observation that the integrand is positive in $\left(0,\frac{\epsilon}{2}\right)$ and negative in $\left(\frac{\epsilon}{2},\infty\right)$. Furthermore, we have
\begin{align}
\int_{0}^{\infty}d\Delta^\prime f(\Delta^\prime)=0
\end{align}
Using the above facts, one can always choose $0<\epsilon_1<\frac{\epsilon}{2}$ and $\frac{\epsilon}{2}<\epsilon_2<\infty$ such that 
\begin{align}\label{1}
\int_{0}^{\epsilon_1}d\Delta^\prime\ f(\Delta^\prime)&= -\int_{\epsilon_2}^{\infty}d\Delta^\prime\ f(\Delta^\prime)\\
\label{2} \int_{\epsilon_1}^{\epsilon_2}d\Delta^\prime\ f(\Delta^\prime)&=0
\end{align}
This is basically guaranteed by the continuity. We choose $\epsilon_1$ such that  $0<\epsilon_1<\frac{\epsilon}{2}$ and consider the function $F(y)=\int_{\epsilon_1}^{y} dx\ f(x)$. Now $F(y)$ is a continuous function. It is positive when $y=\frac{\epsilon}{2}$ and negative when $y\to\infty$. Thus by continuity, there exists $\frac{\epsilon}{2}<\epsilon_2<\infty$ such that the eq.~\eqref{1} holds. The shaded region in the figure.~\ref{fig:func} is the area under the function $f$ restricted to the interval $[\epsilon_1,\epsilon_2]$ so that the eq.~\eqref{1} is satisfied.

 Now we note that 
\begin{align}\label{2}
\int_{\epsilon_1}^{\epsilon_2}dx\ f(x) \exp\left[\frac{-x^2}{2\pi\sqrt{\frac{c}{3}}\Delta^{\frac{3}{2}} }\right]\ge 0
\end{align}
and 
\begin{align}\label{3}
\int_{0}^{\epsilon_1}dx\ f(x) \exp\left[\frac{-x^2}{2\pi\sqrt{\frac{c}{3}}\Delta^{\frac{3}{2}} }\right] &\ge\exp\left[\frac{-\epsilon_1^2}{2\pi\sqrt{\frac{c}{3}}\Delta^{\frac{3}{2}} }\right] \int_{0}^{\epsilon_1}dx\ f(x) \\
\label{4}
\int_{\epsilon_2}^{\infty}dx\ f(x) \exp\left[\frac{-x^2}{2\pi\sqrt{\frac{c}{3}}\Delta^{\frac{3}{2}} }\right] &\ge\exp\left[\frac{-\epsilon_2^2}{2\pi\sqrt{\frac{c}{3}}\Delta^{\frac{3}{2}} }\right]\int_{\epsilon_2}^{\infty}dx\ f(x)
\end{align}
where in the second inequality, we have used negativity of $f(x)$ for $x>\frac{\epsilon}{2}$. Combining the last four equations i.e \eqref{1},\eqref{2},\eqref{3},\eqref{4} we can write
\begin{align}
C\ge \Omega \left(\exp\left[\frac{-\epsilon_1^2}{2\pi\sqrt{\frac{c}{3}}\Delta^{\frac{3}{2}} }\right]-\exp\left[\frac{-\epsilon_2^2}{2\pi\sqrt{\frac{c}{3}}\Delta^{\frac{3}{2}} }\right]\right) \underset{\Delta\to\infty}{\simeq} \frac{\left(\epsilon_2^2-\epsilon_1^2\right) \Omega}{2\pi\sqrt{\frac{c}{3}}\Delta^{\frac{3}{2}}}>0
\end{align}
where $\Omega=\int_{0}^{\epsilon_1}dx\ f(x)>0$ is an order one positive number. This clearly proves that as long as $\epsilon>1$, we can neglect the second piece i.e. contributions from the heavy states due to their subleading nature. In fact, one can do much better and show that\footnote{We thank Alexander Zhiboedov for pointing this out in an email exchange.} $C$ falls like $\Delta^{-3/4}$ by noting the following: 
\begin{align}
\nonumber C&=\frac{\epsilon\pi}{8} \exp \left[-\frac{1}{8\pi\sqrt{\frac{c}{3}} \Delta ^{3/2}}\right] \text{Erfi}\left(\frac{1}{2 \sqrt{2\pi\sqrt{\frac{c}{3}}}\Delta ^{3/4}}\right)\\
&-\frac{\epsilon\pi}{8}  e^{-\frac{\sqrt{3}}{8 \pi  \sqrt{c} \Delta ^{3/2}}} \text{Im}\left[\text{Erf}\left(\frac{\sqrt{\frac{\pi }{2}} \left(2 \pi +\frac{i \sqrt{3}}{2 \pi  \sqrt{c} \Delta ^{3/2}}\right)}{\sqrt[4]{3} \sqrt{\frac{1}{\sqrt{c} \Delta ^{3/2}}}}\right)\right]
& \underset{\Delta\to\infty}{\simeq} \frac{\epsilon}{8} \left(\frac{3}{64c}\right)^{1/4}\Delta^{-3/4}\,.
\end{align} 

To summarize, we have proved that for sufficiently large $\Delta$, 
\begin{align}
\int_{\Delta-\frac{\epsilon}{2}}^{\Delta+\frac{\epsilon}{2}}d\Delta^\prime\ \rho(\Delta^\prime) \ge C\rho_0(\Delta)>0
\end{align}

Therefore we have been able to show that the asymptotic gap between two consecutive operators is bounded above by $\epsilon$, where $\epsilon>1$. Now one can choose $\epsilon$ to be arbitrarily close to $1$, which proves that the optimal bound is exactly $1$. The analysis can be carried over to the case for Virasoro primaries, as pointed out in \cite{Baur}.  This implies that the asymptotic gap between two consecutive Virasoro primaries is bounded above by $1$, which thereby proves the conjecture made in \cite{Baur}. 

{\paragraph{Alternate proof:}
This proof starts from considering the following function for $\sigma \in (0,1)$: 
\begin{equation}\label{def:func}
\begin{aligned}
\phi_-(\Delta')=\frac{\sigma}{1-\left(\frac{\Delta^\prime-\Delta}{\epsilon}\right)^{2}}\left(\frac{\sin\left(\frac{\pi(\Delta^\prime-\Delta)}{\epsilon}\right)}{\frac{\pi(\Delta^\prime-\Delta)}{\epsilon}}\right)^{2}+\frac{(1-\sigma)\cos^{2}\left(\frac{\pi\left(\Delta^\prime-\Delta\right)}{\epsilon}\right)}{1-4\left(\frac{\Delta^\prime-\Delta}{\epsilon}\right)^2}
\end{aligned}
\end{equation}

It can be verified that for $\epsilon \to 1^{+}$ and $\sigma \to 0^{+}$, $c_+$ approaches to $0$ from the positive side and the $\text{max}\left\{x: \phi_-(\Delta+x)>0\right\}$ approaches to $0.5$ adn $x<0.5$ the function remains positive. This also proves the conjecture. In fact, one can see that for $\epsilon=1$, $c_-$, as defined in eq.~\eqref{def:cpm} approaches to $0.5$ as $\sigma\to 1^{-}$, starting from $c_-=0$ for $\sigma=0$, as shown in figure~[\ref{fig:newproof}].
\begin{figure}[h!]
 \centering
\includegraphics[scale=0.7]{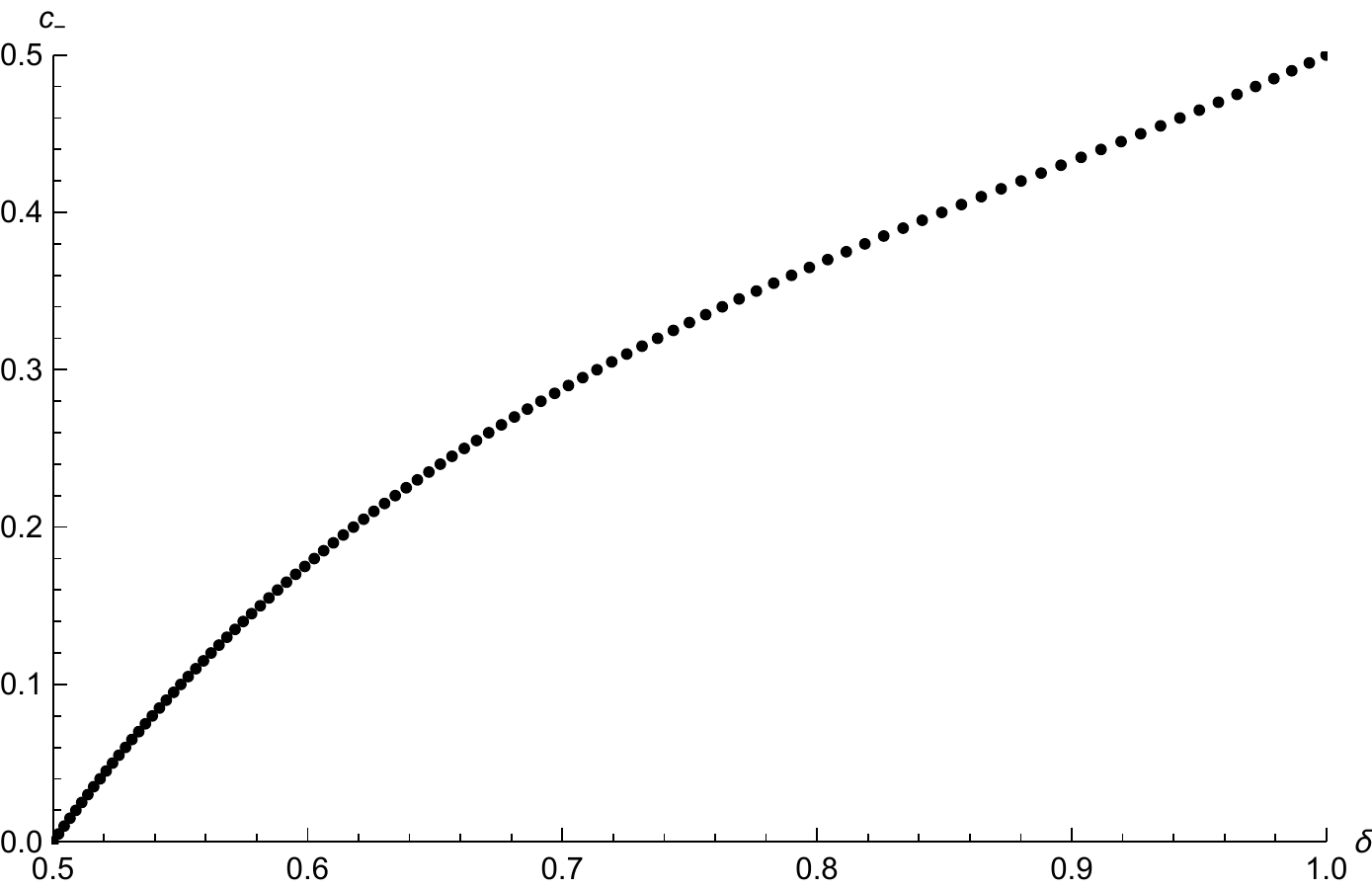}
\caption{$\mathrm{Exp}[s_{\pm}]\equiv c_-\ $ as a function of $\delta$, the half-width of the energy window, obtained from the function defined in eq.~\eqref{def:func}.} 
\label{fig:newproof}
\end{figure}
This observation regarding the behavior of $c_-$ around $\delta=1$ is related to what  is depicted in figure~\ref{fig:snew}.}

\section{Brief discussion}\label{conc}
{In this work, we have improved the existing upper and lower bound on the $O(1)$ correction to the density of states in $2$D CFT at high energy. Since one cannot theoretically deduce a universal correction suppressed compared to $O(1)$ correction, any improvement in the bound on $O(1)$ correction is meaningful. Furthermore, we have shown the optimality of the lower bound in the limit $\delta\to 1^{-}$. We have also proven the conjectured upper bound on the gap between Virasoro primaries. In particular, we have shown that there always exists a Virasoro primary in the energy window of width greater than $1$ at large $\Delta$. This is a quantitative universal measure of/bound on how sparse a CFT spectrum can be asymptotically. In particular, this rules out the possibility of having CFTs with Hadamard like gap. We have also unraveled a curious connection between the sphere packing problem and the problem of finding the lower bound using bandlimited functions. This puts an upper bound on the lower bound for $\delta\leq 1$ using bandlimited functions. As mentioned, for $\delta=1$, the best bound is achievable and in fact the optimal one owing to saturation by Monster CFT.}\\

We have provided a systematic way to estimate how tight the bound can be made using bandlimited functions. Since there is still a gap between the achievable bound and the bound on the bound, there is scope for further improvement (except in the limit $\delta\to 1^{-}$, where we have shown optimality). Ideally, one would like to close this gap, which might be possible either by sampling more points and leveraging the positive definiteness condition on a bigger matrix, or by choosing some suitable function which would make the achievable bound closer to the bound on the bound. Another possible way to obtain the bound on bound is to use a known $2$D CFT partition functions, for example $2$D Ising model and explicitly evaluate $s(\delta,\Delta)$. It would be interesting to see how the bound on bound obtained in this paper compares to the one which can be obtained from the $2$D Ising model. For example, one can verify that the bound on bound obtained here is stronger than that could be obtained from $2$D Ising model\footnote{We thank Alexander Zhiboedov for raising this question of how our bound compares to $s(1.7,\Delta)$ for the $2$D Ising model, as found in \cite{Baur}.} for $\delta=1$. In fact, this happens to be case for different other values of $\delta$ as well. It would be interesting to explore this further analytically.\\

The utility of the technique developed here lies beyond the $O(1)$ correction to the Cardy formula. We expect the technique to be useful whenever one wants to leverage the complex Tauberian theorems, for example in \cite{Pal:2019zzr,Pal:2019yhz,Collier:2019weq}. As emphasized in \cite{Baur}, the importance of Tauberian theorems lies beyond the discussion of $2$D CFT partition functions, especially in investigating Eigenstate Thermalization Hypothesis \cite{deutsch1991quantum,srednicki1994chaos,PhysRevX.8.021026,rigol2008thermalization} in $2$D CFTs\cite{Lashkari:2016vgj,Brehm:2019fyy,Brehm:2018ipf,Basu:2017kzo,Maloney:2018yrz,Maloney:2018hdg,Dymarsky:2018iwx,Dymarsky:2019etq,Datta:2019jeo,Kusuki:2019gjs,Hikida:2018khg,Romero-Bermudez:2018dim}. It is worth mentioning that the use of extremal functionals provided us with sharper inequalities in CFT \cite{Mazac:2016qev,Mazac:2018mdx,Mazac:2018ycv,Hartman:2019pcd}. One can hope to blend the Tauberian techniques with techniques involving extremal functionals to investigate the landscape of $2$D CFT more. We end with a cautious remark that if we relax the condition of using bandlimited functions, the bound on bounds would not be applicable and it might be possible to obtain nicer achievable bounds on the $O(1)$ correction to the Cardy formula. {Nonetheless, we emphasize that the lower bound for $\delta\to1^{-}$ is indeed optimal, thus not restricting to bandlimited functions would not provide us with anything more. }

\section*{Acknowledgements}
The authors thank Denny's for staying open throughout the night.  The authors acknowledge helpful comments and suggestions from Tom Hartman, Baur Mukhametzhanov, and especially, Alexander Zhiboedov. The authors thank Shouvik Datta for encouragement. SG wishes to acknowledge Pinar Sen for some fruitful and illuminating discussions on what makes the Fourier transform of a function positive, which helped him arrive at the answer by thinking about autocorrelation functions and power spectral densities. SP acknowledges a debt of gratitude towards Ken Intriligator and John McGreevy for fruitful discussions and encouragement. SP thanks Shouvik Datta and Diptarka Das for introducing him to the rich literature of CFT in 2017. This work was in part supported by the US Department of Energy (DOE) under cooperative research agreement DE-SC0009919 and Simons Foundation award \#568420. SP also acknowledges the support from Inamori Fellowship and Ambrose Monell Foundation and DOE grant DE-SC0009988.

{\bibliographystyle{bibstyle2017}
\bibliography{refs}
}

\end{document}